\documentclass[conference]{IEEEtran}
\usepackage{cite}
\usepackage{amsmath,amssymb,amsfonts}
\usepackage{algorithmic}
\usepackage{graphicx}
\usepackage{textcomp}
\usepackage{xcolor}
\usepackage{fancyhdr}
\usepackage[hyphens]{url}

\def\BibTeX{{\rm B\kern-.05em{\sc i\kern-.025em b}\kern-.08em
    T\kern-.1667em\lower.7ex\hbox{E}\kern-.125emX}}

\usepackage{microtype}
\usepackage{graphicx}
\usepackage{booktabs} 

\usepackage{amsmath}
\usepackage{algorithmic}
\usepackage[linesnumbered]{algorithm2e}
\usepackage{color}
\usepackage[T1]{fontenc}

\usepackage{makecell}

\usepackage{breakurl}

\usepackage{graphicx}

\usepackage{setspace}
\usepackage{comment}
\usepackage{subfigure}
\usepackage{caption}
\usepackage[most]{tcolorbox}

\usepackage{endnotes,microtype,xspace,graphicx,fancyvrb,multirow}

\usepackage{paralist}

\usepackage{mathptmx} 

\newcommand{\ignore}[1]{}
\usepackage{fancyhdr}
\usepackage[normalem]{ulem}
\usepackage[hyphens]{url}
\usepackage{longtable}
\usepackage{microtype}
\usepackage{graphicx}
\usepackage{caption}
\usepackage{bbding}

\usepackage{tikz}

\usepackage[bookmarks=true,breaklinks=true,letterpaper=true,colorlinks,linkcolor=blue,citecolor=blue,urlcolor=red]{hyperref}
\usepackage[font=small]{caption}
\newcommand{\PCignore}[1]{}

\def\Snospace~{\S{}}

\newcommand{\insertFigure}[2]{
    \begin{figure}[t]
\setlength{\abovecaptionskip}{-1pt}
\setlength{\belowcaptionskip}{-1pt}
        \centering
        \includegraphics[width=\linewidth]{figs/#1.pdf}
        \caption{\small #2}
        \label{fig:#1}
    \end{figure}
}
\newcommand{\insertFigureNew}[3]{
    \begin{figure}[h]
\setlength{\abovecaptionskip}{-1pt}
\setlength{\belowcaptionskip}{-1pt}
        \centering
        \includegraphics[width=#3\linewidth]{figs/#1.pdf}
        \caption{\small #2}
        \label{fig:#1}
    \end{figure}
}

\newcommand{\insertWidePng}[2]{

    \begin{figure*}[h]
    \setlength{\abovecaptionskip}{-1pt}
    \setlength{\belowcaptionskip}{-4pt}
        \centering
        \includegraphics[width=\textwidth]{figs/#1.png}
	    \vspace{-2mm}
        \caption{\small #2}
    	\vspace{-2mm}
        \label{fig:#1}
    \end{figure*}
}

\newcommand{\insertWideFigureNew}[3]{

    \begin{figure*}[ht]
    \setlength{\abovecaptionskip}{-1pt}
    \setlength{\belowcaptionskip}{-4pt}
        \centering
        \includegraphics[width=#3\linewidth]{figs/#1.pdf}
	    \vspace{-2mm}
	    \bigskip
        \caption{\small #2}
        \bigskip
    	\vspace{-4mm}
        \label{fig:#1}
    \end{figure*}
}

\newcommand{\squishlist}{
 \begin{list}{$\bullet$}
  { \setlength{\itemsep}{0pt}
     \setlength{\parsep}{3pt}
     \setlength{\topsep}{3pt}
     \setlength{\partopsep}{0pt}
     \setlength{\leftmargin}{1.5em}
     \setlength{\labelwidth}{1em}
     \setlength{\labelsep}{0.5em} } }

\newcommand{\squishlisttwo}{
 \begin{list}{$\bullet$}
  { \setlength{\itemsep}{0pt}
     \setlength{\parsep}{0pt}
    \setlength{\topsep}{0pt}
    \setlength{\partopsep}{0pt}
    \setlength{\leftmargin}{2em}
    \setlength{\labelwidth}{1.5em}
    \setlength{\labelsep}{0.5em} } }

\newcommand{\squishend}{
  \end{list}  }

\newcommand{\TK}[1]{\textcolor{blue}{TK: #1}}

\newcommand{\ncu}{ACE\xspace}
\usepackage{graphicx}

\title{Enabling Compute-Communication Overlap in Distributed Deep Learning Training Platforms\vspace{-27pt}} 
\author{}

\begin{document}
\author{\IEEEauthorblockN{\\Saeed Rashidi\IEEEauthorrefmark{1}, Matthew Denton\IEEEauthorrefmark{1},
Srinivas Sridharan\IEEEauthorrefmark{2}, Sudarshan Srinivasan\IEEEauthorrefmark{3}, \\Amoghavarsha Suresh\IEEEauthorrefmark{4}, Jade Nie\IEEEauthorrefmark{2}, and
Tushar  Krishna\IEEEauthorrefmark{1}}
\IEEEauthorblockA{\IEEEauthorrefmark{1}Georgia Institute of Technology, 
Atlanta, USA\\
}
\IEEEauthorblockA{\IEEEauthorrefmark{2}Facebook, 
Menlo Park, USA\\}
\IEEEauthorblockA{\IEEEauthorrefmark{3}Intel, 
Bangalore, India\\}
\IEEEauthorblockA{\IEEEauthorrefmark{4}Stony Brook University, 
Stony Brook, USA\\
\textit{ saeed.rashidi@gatech.edu, 
ssrinivas@fb.com, sudarshan.srinivasan@intel.com, tushar@ece.gatech.edu}}}

\maketitle
\pagestyle{plain}

\begin{abstract}
Deep Learning (DL) training platforms are 
built 
by interconnecting multiple DL accelerators (e.g., GPU/TPU) via fast, customized interconnects 
with 100s of gigabytes (GBs) of 
bandwidth.
However, as we identify in this work,
driving this bandwidth is quite challenging. This is because there is a pernicious balance between using the 
accelerator's 
compute and memory  
for both DL computations and communication.

This work makes two key contributions.
First, via real system measurements and detailed modeling, we provide an understanding of compute and memory bandwidth demands 
for DL compute and comms.
Second, 
we propose a novel
DL collective communication accelerator 
called \emph{Accelerator Collectives Engine} (ACE) 
that sits alongside the compute and networking engines at the accelerator endpoint.
\ncu frees up the endpoint's 
compute and memory resources
for DL compute, 
which in turn
reduces the required memory BW by 3.5$\times$ on average to drive the same network BW compared to state-of-the-art baselines. For modern DL workloads and different network sizes, \ncu, on average,  increases the effective network bandwidth utilization by 1.44$\times$ (up to 2.67$\times$), resulting in an average of 1.41$\times$ (up to 1.51$\times$), 1.12$\times$ (up to 1.17$\times$), and 1.13$\times$ (up to 1.19$\times$) speedup in iteration time for ResNet-50, GNMT and DLRM when compared to the best baseline configuration, respectively. 
\end{abstract}
\section{Introduction}\label{sec:intro}

Deep Learning (DL) and Deep Neural network (DNN) models are being deployed pervasively across a wide range of real-world application domains~\cite{Resnet, gnmt, dlrm}.
The size and computational requirements of these DNN models are growing at an unparalleled rate, 2$\times$ every 3.4 months \cite{OPENAI}, to handle the unrelenting growth in data and workload requirements. 
The advent of energy-efficient 
accelerators capable of handling these large models and the need for accelerating training time when dealing with 10s to 100s of petabytes of input data is raising the demand for faster and more efficient DL \textit{training} solutions. This can only be achieved through scalable and efficient \textit{distributed} training since a single accelerator cannot satisfy the compute, memory, and I/O requirements of today's state-of-the-art DNNs. 

\begin{figure}[h]
\begin{center}
\includegraphics[scale=0.255]{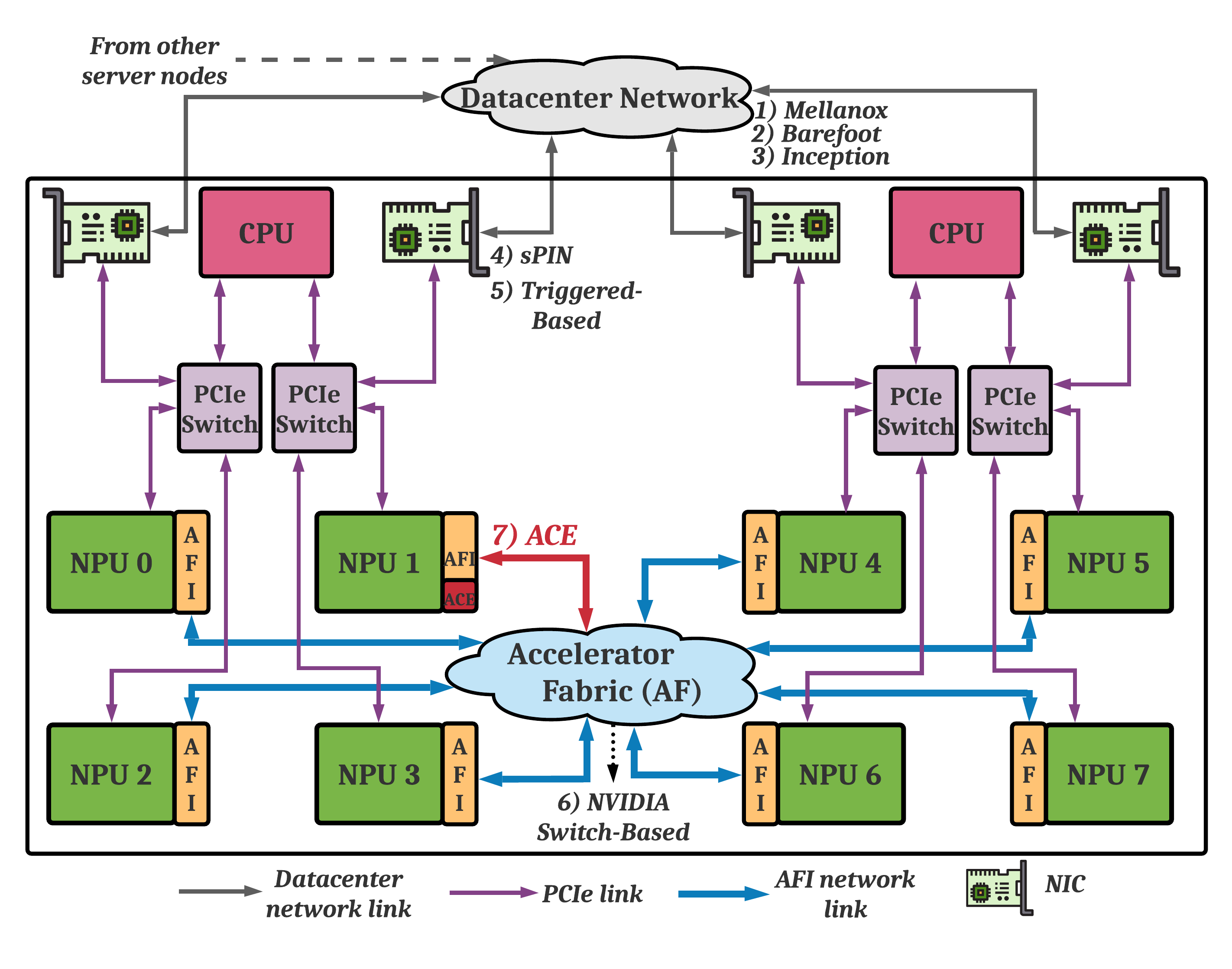}
\end{center}
\vspace{-2mm}
\caption{Architecture of a DL Training Platform (e.g., Google Cloud TPU , NVIDIA DGX-1/DGX-2, Facebook Zion, Intel Xe-GPU \cite{IntelGPU}). Accelerator Fabrics may be point-to-point (e.g, Hypercube Mesh in Zion \cite{FBLargeScaleTraining,Zion}, Torus in TPU \cite{TPU}, Habana \cite{HabanaPtP}) or switch-based (e.g., NVswitch in DGX-2  \cite{dgx2}). The numbers indicate prior work on offloading collectives and contrasted in 
\autoref{table:related_offload}}
\vspace{-2mm}
\label{fig:overviewOfServer}
\end{figure}

\insertFigure{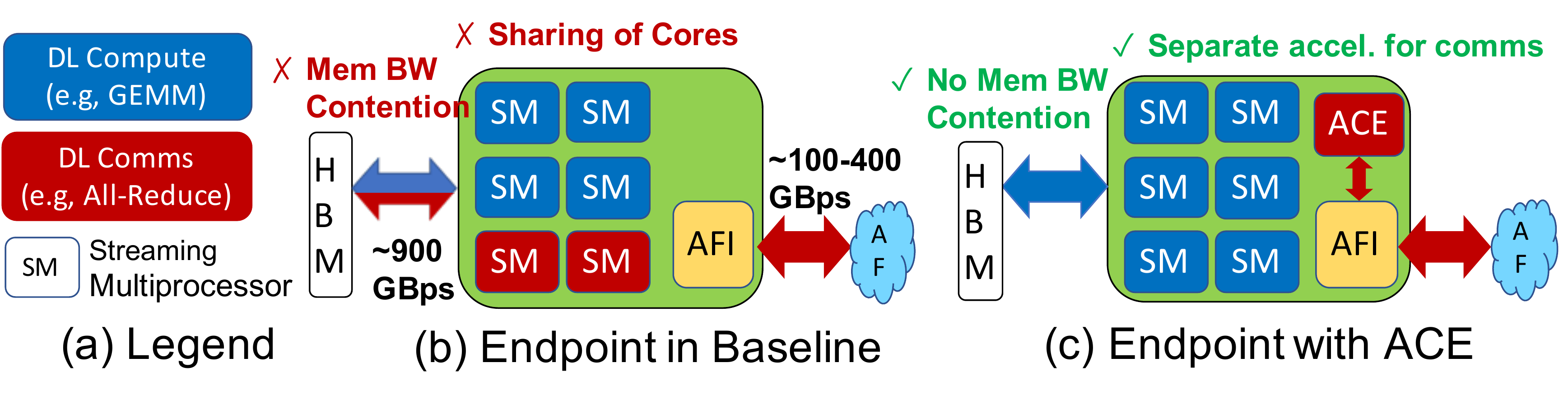}{Endpoint NPU Node in baseline systems (today) and with ACE (proposed). In baseline, collective communication tasks contend for NPU cores (e.g., SMs in a GPU) and memory bandwidth.}


The need for distributed training has led to the emergence of \textit{DL Training Platforms} such as Google Cloud TPU~\cite{cloudtpu}, Facebook Zion~\cite{Zion}, NVIDIA DGX-1~\cite{dgx1}/DGX-2~\cite{dgx2}, and Intel Xe-GPU \cite{IntelGPU}. 
A common feature in all these platforms is the 
use of a customized interconnect connecting the 
various accelerators together 
(in addition to the traditional cache-coherent shared-memory CPU network such as QPI and 
the TCP-IP/RoCE/Infiniband based
datacenter network accessible via PCIe/Infiniband NICs), as shown in \autoref{fig:overviewOfServer}. 
In this paper, we use the term \emph{Neural Processing Units} (NPUs) to describe the accelerator (GPUs, TPUs, FPGAs) and the term \emph{Accelerator Fabric} (AF) 
for the accelerator network\footnote{There is no consistent terminology yet in the community for this fabric. Facebook Zion~\cite{Zion} uses the same terminology as ours, Google Cloud TPU~\cite{cloudtpu} calls it the Inter-core Interconnect (ICI), some papers have called it the device-side interconnect~\cite{mem_centric} and shared memory fabric~\cite{InNetwork}.}.

The AF is different from traditional NIC-based 
datacenter networks for two reasons: (i) NPUs can directly talk to each other without CPU/NIC intervention, (ii) the AF provides 10's of times higher bandwidth  
than the datacenter 
network (e.g. 12.5 GB/s vs. 500 GB/s) as it 
employs a mix of on-package interconnects to bind NPU chiplets together~\cite{mcm-gpu,IntelGPU} 
and custom high-bandwidth interconnects like Xe-Link~\cite{IntelXe} or NVlink~\cite{NVlink} to connect these packages together.

While there has been significant 
work from the architecture community 
on accelerating the 
compute portion of DL training via 
efficient accelerators~\cite{TPU,MAERI,maestro} and memory systems~\cite{RecNMP,TensorDIMM}, 
an often ignored part of the story is communication.
Distributed DL training fundamentally involves splitting the DNN model, training data, or both across multiple NPUs. These schemes are referred to as model, data, and hybrid parallel, respectively.
The parallelization strategy in turn dictates the communication required between NPUs. 
This communication is \textit{collective} in nature, i.e., all NPUs synchronize input/output/error activations during the forward/backward pass and gradients during the backward pass.
Specifically, two collective operations: 
all-to-all and all-reduce, 
occur heavily during distributed DL training. These operations are often latency-sensitive (since the next layer of the DNN cannot proceed until gradients have been synchronized\footnote{We assume  synchronous updates which provide better guarantees at convergence than asynchronous updates.
}) and bandwidth-hungry due to 
the large sizes of the activations/gradients and limited network bandwidth 
and hence can easily become bottleneck  \cite{CommBottleneck1,CommBottleneck2}.

One of the primary techniques employed today to minimize the impact of communication is to overlap it behind compute. This is enabled via
clever fabric 
topologies~\cite{cloudtpu, dong2020eflops} accompanied by topology-aware collective communication algorithms (e.g., Ring and Double-binary tree for All-reduce) implementations in 
libraries like Intel's MLSL~\cite{mlsl}/oneCCL \cite{oneCCL} and NVIDIA's NCCL~\cite{nccl}. The Cloud TPU in fact boasts of contention-free O(1)
scalability of All-reduce on their Torus~\cite{gshard}.
However, in this work, we identify that getting perfect compute-comms overlap on training platforms, 
despite using 
optimized topologies and collective algorithm implementations,
is prohibitive due to a key architectural 
bottleneck at the endpoint; both compute and communication 
contend for the same shared resources: compute units and memory bandwidth.

\autoref{fig:baseline_vs_ace} shows an NPU and its connected peripherals. 
We indicate two sources of inefficiency.
(i) \textbf{Compute:} a portion of the compute resources of the NPU is used for performing collective updates and driving the network, which reduces the compute available to training computations (e.g., GEMMs during the forward and backward passes),
(ii) \textbf{Memory Bandwidth:} the received gradients need to be written to and read from memory, in turn reducing the bandwidth available for the actual training computations which are known to be highly bandwidth-intensive in modern DNNs such as recommendations~\cite{dlrm} and NLP~\cite{deepspeed}.
This problem gets exacerbated as the recent advances in the multi-chip module (MCM) packaging technologies (i.e., silicon interposer) and high-bandwidth inter-package networks provide enormous network BW to the NPUs at the AF level, thus making it more challenging for the NPUs to fully drive the network.

We corroborate the aforementioned issues via real system measurements on a DGX-2 system running both microbenchmarks, real Deep Learning Recommendation Model (DLRM)~\cite{dlrm}, and Megatron-LM~\cite{Megatron} workloads (\autoref{sec:motivation}).
The implication of these issues is the under-utilization of the available AF bandwidth as systems scale, 
which we demonstrate via a detailed simulator.
\textbf{\textit{To the best of our knowledge, this is the first work to identify these issues.}}



To address these inefficiencies, we propose
\emph{Accelerator Collectives Engine (ACE)}, which is an accelerator  
for DL training collectives. \ncu sits alongside the \emph{Accelerator Fabric Interface} (AFI) \cite{Zion} that interfaces NPU to the AF network and handles the communication protocol.  
\ncu houses compute units for running
a variety of collective 
algorithms and scratchpads to cache gradients.
This in turn frees up NPU resources and memory bandwidth for the GEMM computations, as shown in \autoref{fig:baseline_vs_ace}.
\autoref{table:related_offload} shows the difference between the previous works on communication offload in datacenter networks vs. \ncu 
that is also visualized in \autoref{fig:overviewOfServer}. To the best of our knowledge, \textbf{\textit{this is the first endpoint-based offload scheme that is tuned for distributed training for the AF}}.

With ACE, we demonstrate that the compute speeds up and the available AF bandwidth can be utilized much more efficiently, decreasing overall training time.

\begin{table}[!t]
\caption{Comparison of previous SmartNIC and switch offload schemes against \ncu. Coll. => Collective, DCN => Datacenter Network, AF => Accelerator Fabric} 
\footnotesize
 \label{table:related_offload}

\begin{tabular}{|p{3.3em}|p{2.2em}|p{2.7em}|p{3.7em}|p{3.2em}|p{2.6em}|p{3.0em}|}
\hline
\textbf{Scheme} & \textbf{App} & \textbf{Offload} & \textbf{Protocol} & \textbf{Topology} & \textbf{Aggr.} & \textbf{Network}  \\
\hline
Mellanox \cite{Mellanox} & HPC, DL & Switch & Infiniband & Switch-based & Coll. & DCN \\\hline
Barefoot \cite{Barefoot} & DL & Switch & Ethernet & Switch-based & Coll. & DCN \\\hline
sPIN \cite{NIC2} & HPC & NIC & Ethernet & Switch-based & Coll. & DCN \\\hline
Trigger-ed~\cite{NIC1} & HPC &  NIC & RDMA-based & Flexible & Coll. & DCN \\\hline
Incept-ionn~\cite{Hadi1} & DL &  NIC & Ethernet & Tree & Param Server & DCN \\\hline
NVIDIA \cite{InNetwork} & DL & Switch & Shared Memory & Switch-based & Coll. & AF \\\hline
ACE & DL & Endpoint & RDMA-based & PtToPt, Switch & Coll. & AF \\\hline
\end{tabular}
\end{table}

\insertWidePng{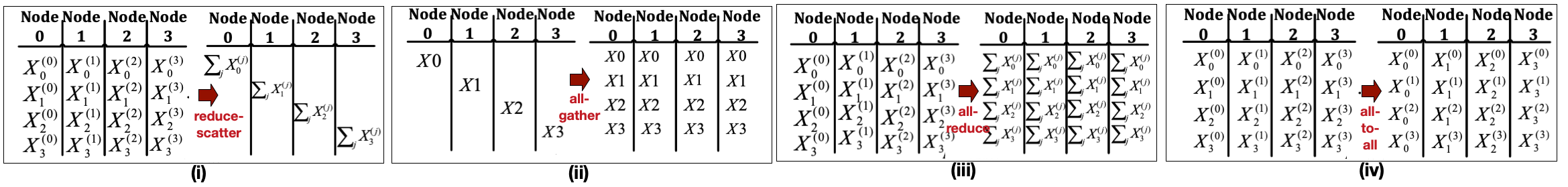}{Overview of collective communication operations used in DNN training networks.}

This paper makes the following contributions:
\squishlist
    \item We identify a set of key challenges in the end-points of DL training platforms related to compute and memory bandwidth 
    that can limit the utilization of available AF bandwidth for future training platforms (\autoref{sec:motivation}). 
    \item We propose a novel microarchitecture called \ncu designed to handle collective communication and efficiently enhance AF network utilization (\autoref{sec:design}).  ACE  frees  up  the  endpoint’s  compute and  memory  resources  for  DL  compute,  which  in  turn  reduces the required memory BW by 3.5× on average to drive the same network  BW  compared  to  state-of-the-art  baselines. On average, \ncu increases the effective network bandwidth utilization by 1.44$\times$ (up to 2.67$\times$), resulting in average of 1.41$\times$ (up to 1.51$\times$), 1.12$\times$ (up to 1.17$\times$), and 1.13$\times$ (up to 1.19$\times$) speedup in iteration time for ResNet-50, GNMT and DLRM when compared to the best baseline configuration, respectively.
    \item Finally, we show that the reduction in the memory BW requirement (enabled by \ncu) allows for performing better workload-specific optimizations, with the specific example of optimizing DLRM.
\squishend

The rest of the paper is organized as follows: 
\autoref{sec:background} presents the necessary background for distributed training systems. \autoref{sec:motivation} establishes the challenges in scaling DL training, specifically focused on critical bottlenecks in the endpoint that inhibit efficient network utilization. \autoref{sec:design} describes the ACE microarchitecture. This is followed by a detailed description of our evaluation and simulation methodology in \autoref{sec:methodology} and the experimental results in \autoref{sec:results}. Next, we compare our work against the related works in \autoref{sec:related}. Finally, we conclude the paper in \autoref{sec:conclusions}. 


\vspace{-2mm}
\section{Background} \label{sec:background}
Training DNNs involves iteratively refining the parameters (aka weights) of the network by solving a non-convex, non-linear optimization problem to minimize a loss function.
Here, we provide background on distributed training~\cite{ouyang2020communication, DBLP:journals/corr/0002AMVSKKD16}.

\textbf{Parallelization.} 
The most common parallelization technique for speeding 
up DL training is called \emph{data parallelism}. It 
replicates the entire model on multiple nodes to take advantage of the large number of input samples. Every node computes partially trained weight gradients for its subset of samples of the input data, aka mini-batch. At the end of each iteration, nodes exchange their partially trained weight gradients and perform the SGD operation to update the weights gradients accumulated from all nodes. The updated weights, are then used in the forward pass of the next iteration. 
In \emph{model parallelism}, all nodes have the same datasets and work on the same mini-batch, but the model is divided among nodes. Each node thus produces a part of the output activations and input gradients during the forward pass and back-propagation, respectively, and these values must be communicated across all nodes  to enable forward pass and back-propagation.


\textbf{Collective Communication Operations.}
Exchange of input/weight gradients and output activations among the nodes, depending on the parallelism approach, is known as "collective communication". 
In general, four different collective communication operations are the main contributor in DNN training communication~\cite{astrasim, InNetwork}, as shown in \autoref{fig:collectives_table}: 
(i) reduce-scatter, (ii) all-gather, (iii) all-reduce, and (iv) all-to-all.
Reduce-scatter reduces (e.g. sum) all data, initially residing in the nodes, such that at the end each node has a portion globally reduced data. All-gather gathers the data, initially scattered across nodes, such that at the end, all of the nodes have all of the data. All-reduce can be thought of as a reduce-scatter followed by an all-gather. In all-to-all, each node needs to send a different portion of data to other nodes. All-reduce is the dominant communication pattern observed in the DL training for exchanging gradients and activations in various parallelism schemes. However, all-to-all is used in some scenarios such as table embedding exchanges for recommendation models such as Facebook DLRM~\cite{dlrm}.

\textbf{Topology-Aware Collective Algorithm.}
Collectives have efficient implementation algorithms based on the underlying topology.
Libraries like Intel oneCCL~\cite{oneCCL} and NVIDIA's NCCL~\cite{nccl} provide different implementations 
for collectives, 
such as ring-based, 
tree-based, 
hierarchical, direct all-reduce/all-to-all,
to optimize for the available bandwidth in the underlying topology  \cite{collective1,collective2,collective3}.
We will discuss this further in \autoref{sec:methodology}, where we consider topology-aware collectives for evaluating our target systems.

\insertFigureNew{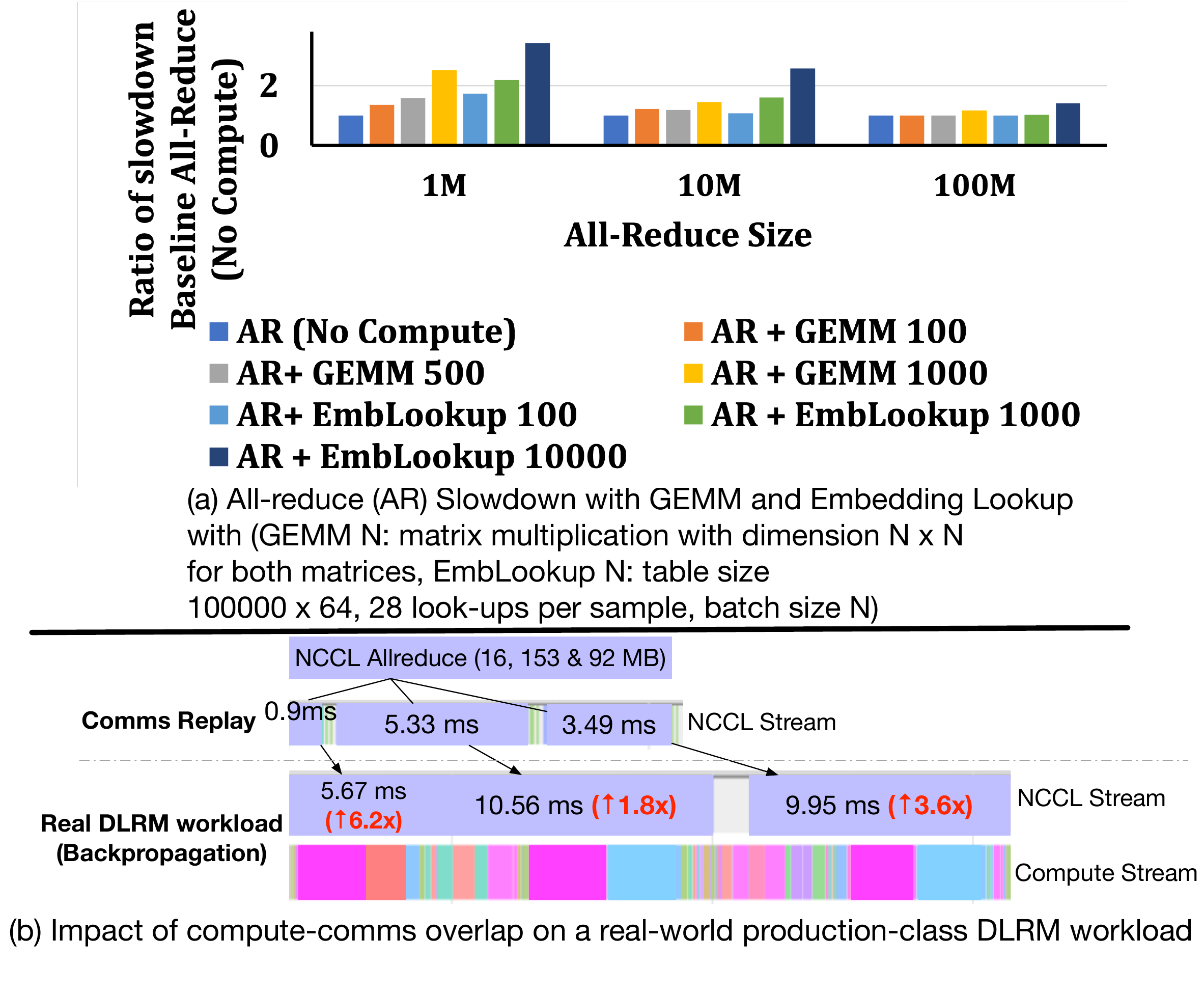}{Slowdown of the communication when overlapped with computation. The platform is NVIDIA V100 8-GPU system connected with NVSwitch offering 150 GB/s available network BW per GPU.}{1}
\insertFigureNew{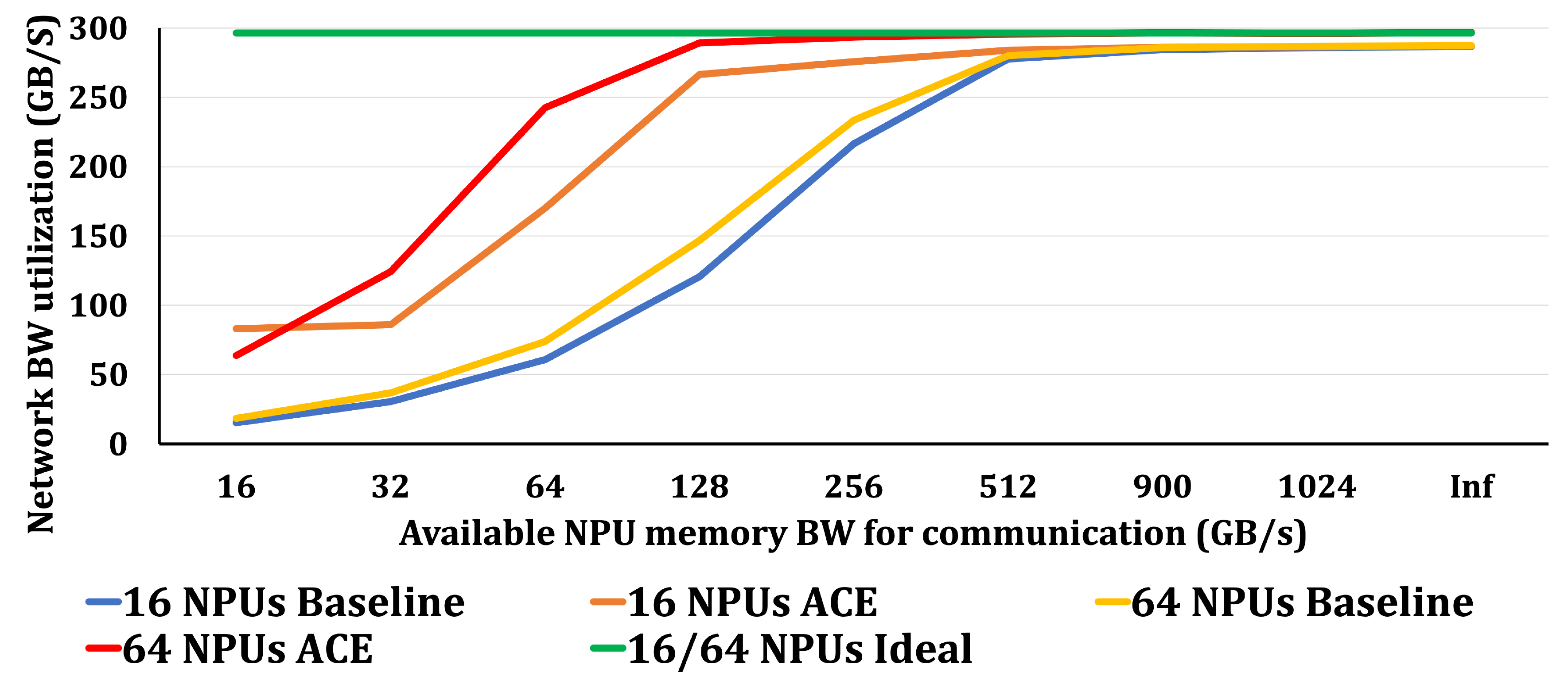}{The sensitivity analysis that shows how the AF network utilization is affected as the amount of NPU memory BW available for DL communication is increased. The communication is a single 64MB all-reduce. Ideal system assumes that received data (during collective communication) is magically  processed and ready only after 1 cycle and is used to get and upper bound for network performance. The baseline system assumes all SMs are available for the communication task, while \ncu does not consume any SM within the NPU.}{0.9}

\insertFigureNew{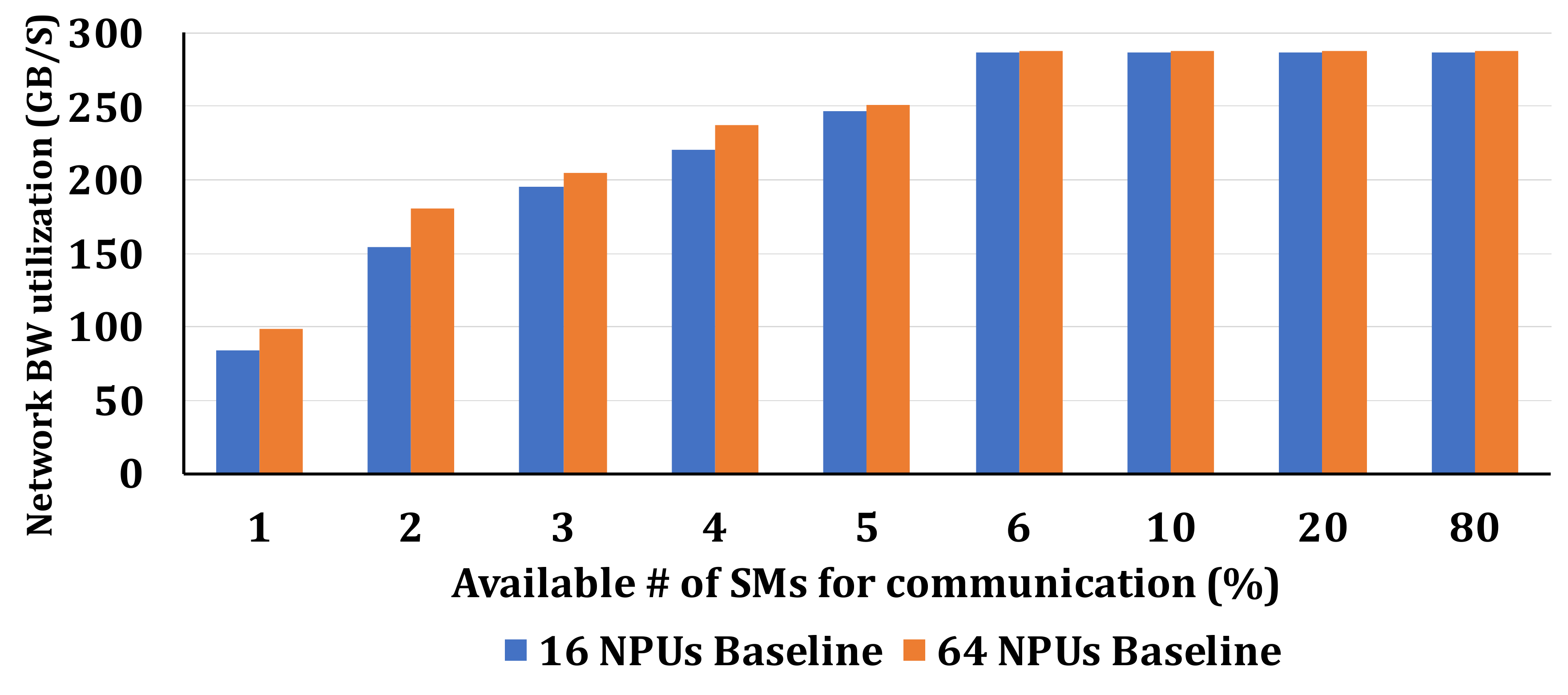}{The sensitivity analysis that shows how the network utilization is affected in the baseline system, as the available \# of SMs for DL communication changes. The communication is a single 64MB all-reduce. It is assumed that all NPU memory BW is available for the communication. Note that \ncu does not rely on NPU SMs for communication and hence, this experiment is not applicable for it.}{0.9}

\section{Motivation: Poor AF BW Utilization}\label{sec:motivation}


In this section, we highlight some of the critical challenges in achieving high network bandwidth utilization for DL training workloads. \autoref{fig:baseline_vs_ace} 
qualitatively summarizes our findings. 


\textbf{NPU Compute Availability:} On modern accelerator systems, a fraction of NPU cores (e.g. CUDA cores) \emph{concurrently} execute kernels pertaining to DL collective communication while the majority of compute cores execute DL computation (e.g. GEMMs and Convolutions). 
Collective operations, such as all-reduce, are parallelized across these cores since a single NPU core cannot fully saturate memory bandwidth and network bandwidth, given various bottlenecks in the NPU core - memory - network data-path \cite{mlsl,nccl}.  Each core iterates over a sequence of send/recv operations to/from peers followed by an optional computation on data that was received with locally available data (e.g. reduction sum). Thus, the optimal number of NPU cores is a function of network bandwidth saturated by a single core, memory bandwidth available to a single core, communication algorithm, message size, and so on. 

\textbf{Memory Bandwidth Availability:} 
Collective communication operations require access to sufficiently high memory bandwidth for performing local reduction sum (streaming operation involving two local memory reads, sum operation, and local write) and remote network write (e.g. RDMA write and synchronization) operations. The memory bandwidth requirements is a function of the actual collective operation and is proportional to network bandwidth. 

\textbf{Real System Measurements and Analysis.}
We highlight the resource contention issue between communication and computation by creating a microbenchmark workload and also including the results of real training worklaods for DLRM model   presented in \autoref{fig:motiv_all} and Megatron-LM model.

The microbenchmark executes a compute operation, followed by posting a communication operation, then repeating the compute operation, and finally waiting for communication to complete. We consider two fundamental compute operations common in recommendation models~\cite{dlrm}: matrix multiplication (GEMM) which consumes GPU compute cores and embedding table lookup (EmbLookup) which mainly consumes GPU memory bandwidth. \autoref{fig:motiv_all}.a presents the slowdown of NCCL all-reduce collective on an NVIDIA V100 system with 8 GPUs interconnected via NVSwitch, with a total of 150 GB/s available network BW. 
The slowdown is proportional to the scale of compute operation and all-reduce with a smaller size is more sensitive. Running 100MB all-reduce concurrently with dimension-1000 GEMM (requires 44.8 warps per Stream Multiprocessor (SM)), which is a typical scenario in training recommendation models, will slow down all-reduce by 1.16x. Overlapping 100 MB all-reduce with EmbLookup with batch size 10000 (uses 429.2 GB/s memory bandwidth) slows it down by 1.42x.

To further highlight the performance degradation, we present the slowdown of NCCL all-reduce operations when run concurrently with compute kernels, such as GEMM and EmbLookup, in a real-world PyTorch DLRM workload with batch size 512~\cite{dlrm} compared to the reproduction of the same communication patterns without computation using PARAM replay benchmark~\cite{param}. As shown in \autoref{fig:motiv_all}.b (run on the same system as the above benchmark), we observe the time of 16 MB all-reduce can be increased from $0.9ms$ (without any overlap) to $5.67ms$ (overlapped with GEMM and embedding lookup kernels), up to 6.2x degradation during the backward propagation. Note that such degradation is consistent for all-reduce operations with different sizes as shown in~\autoref{fig:motiv_all}.b and across different epochs.

While PyTorch uses separate CUDA streams for compute and NCCL operations, the actual overlap of these operations changes dynamically based on the CUDA HW scheduler and pending operations. This negative impact of congestion for compute/memory resources by competing kernels results in poor network BW utilization and variability in execution time.

We also evaluated Megatron-LM \cite{Megatron} Workload on our platform (not shown in \autoref{fig:motiv_all}) and gathered communication time in two scenarios: (i) When communication is overlapped with compute, and (ii) When all communications are explicitly issued at the end of 
back-propagation when computation is finished. This is the no-overlap scenario where the training algorithm needs to make a blocking wait until communication is done. On average, overlapping communication with computation degrades the communication performance by $\approx$1.4$\times$  compared to the non-overlap scenario, further pointing to the problem of compute-communication resource contention.

\textbf{Simulation Results.}
To demonstrate that the network BW drive problem is exacerbated as network BW increases in new systems with higher available network BW (i.e. 500 GB/s), we conducted a simple simulation experiment to show how communication performance is affected as the available memory BW (\autoref{fig:motiv}) or NPU cores (\autoref{fig:motiv2}) available for DL communication tasks are varied.
We highlight that in this paper we assumed GPU-like NPUs that consist of Streaming Processor (SM) units as described in \autoref{sec:methodology}. 

\autoref{fig:motiv} shows the results for 16 and 64 NPUs systems running a single 64 MB all-reduce. The other system setup is the same as the systems described in \autoref{sec:methodology}. As can be seen, the ideal system can reach up to 300 GB/s BW utilization, out of available 500 GB/s, since intra-package links (i.e. silicon interposer) become underutilized due to their imbalance speed compared to inter-package links. \autoref{fig:motiv} also shows that the baseline system needs $\approx$ 450 GB/s memory BW on average to reach 90\% of ideal network BW. The reason is that on average, for each message to send out, there are multiple reads/writes issued to move the data and do the local reduction in the baseline system. However, for \ncu, 128GB/s is enough to reach to the 90\% of ideal BW utilization, resulting in $\approx$3.5$\times$ reduction in memory BW requirement to reach the same performance. The reason for such improvement is further elaborated in \autoref{sec:analyticalBound}.


\autoref{fig:motiv2} shows how many Streaming multiprocessors (SMs) are required to prevent the GPU-like compute cores from being the bottleneck in driving the networks. SMs are used to read data from the main memory and inject it into the network \cite{nccl}. For the frequency of 1245 MHz and read/write BW of 64-bytes/cycle, the memory BW is $\approx$ 80 GB/s per SM (please see \autoref{sec:methodology} for more information). Then, We use \autoref{fig:motiv} to calculate how much network BW can be driven as we increase the number of SMs for the communication. For our simulated platforms, 6 SMs are enough to reach to the 450 GB/s memory BW. This is in line with the percentage of core usage for libraries such as oneCCL \cite{oneCCL} and NCCL \cite{nccl}.

\textbf{Takeaway.}
 Overall, we argue that using NPU cores (e.g., CUDA cores) for executing collective communication 
 increases contention for precious compute cores
 and memory bandwidth on the accelerator node.
 The complex interplay between compute-memory-network not only affects collective performance but also the performance of compute kernels executing concurrently. This in turn increases the effective compute time available for overlapping the communication operations - hiding deep inefficiencies.
These challenges are expected to get exacerbated in future training platforms, 
where better compute accelerators and 
 hierarchical bandwidths (e.g., emerging MCM technologies~\cite{DLCircuit,IntelGPU}) will make it harder to hide 
 communication behind compute.

\begin{table}[h]
\centering
\caption{Trade-off between offload at switch vs. endpoint.}
\label{tab:SwitchVsEndpoint}
\resizebox{\linewidth}{!}{%
\begin{tabular}{|l|l|l|l|l|l|l|}
\hline
\begin{tabular}[c]{@{}l@{}}Offload \\ Scheme\end{tabular} &
  \begin{tabular}[c]{@{}l@{}}Increase \\ Available \\ Memory BW\\ \& Compute\\ for Training\end{tabular} &
  \begin{tabular}[c]{@{}l@{}}Endpoint\\ Congestion\\ Reduction\end{tabular} &
  \begin{tabular}[c]{@{}l@{}}Switch Based\\ Topology\\ Compatibility\end{tabular} &
  \begin{tabular}[c]{@{}l@{}}Point-to-Point\\ Topology\\ Compatibility\end{tabular} &
  \begin{tabular}[c]{@{}l@{}}Hybrid\\ Topology\\ Compatibility\end{tabular} &
  \begin{tabular}[c]{@{}l@{}}Various\\ Collective \\ Algorithm\\ Support\end{tabular} \\ \hline
\begin{tabular}[c]{@{}l@{}}Switch\\ Based\end{tabular} &
  \CheckmarkBold &
  \CheckmarkBold &
  \CheckmarkBold &
   &
  Partially &
   \\ \hline
\begin{tabular}[c]{@{}l@{}}Endpoint\\ Based\end{tabular} &
  \CheckmarkBold &
  \CheckmarkBold &
  \CheckmarkBold &
  \CheckmarkBold &
  \CheckmarkBold &
  \CheckmarkBold \\ \hline
\end{tabular}%
}
\end{table}

\insertFigureNew{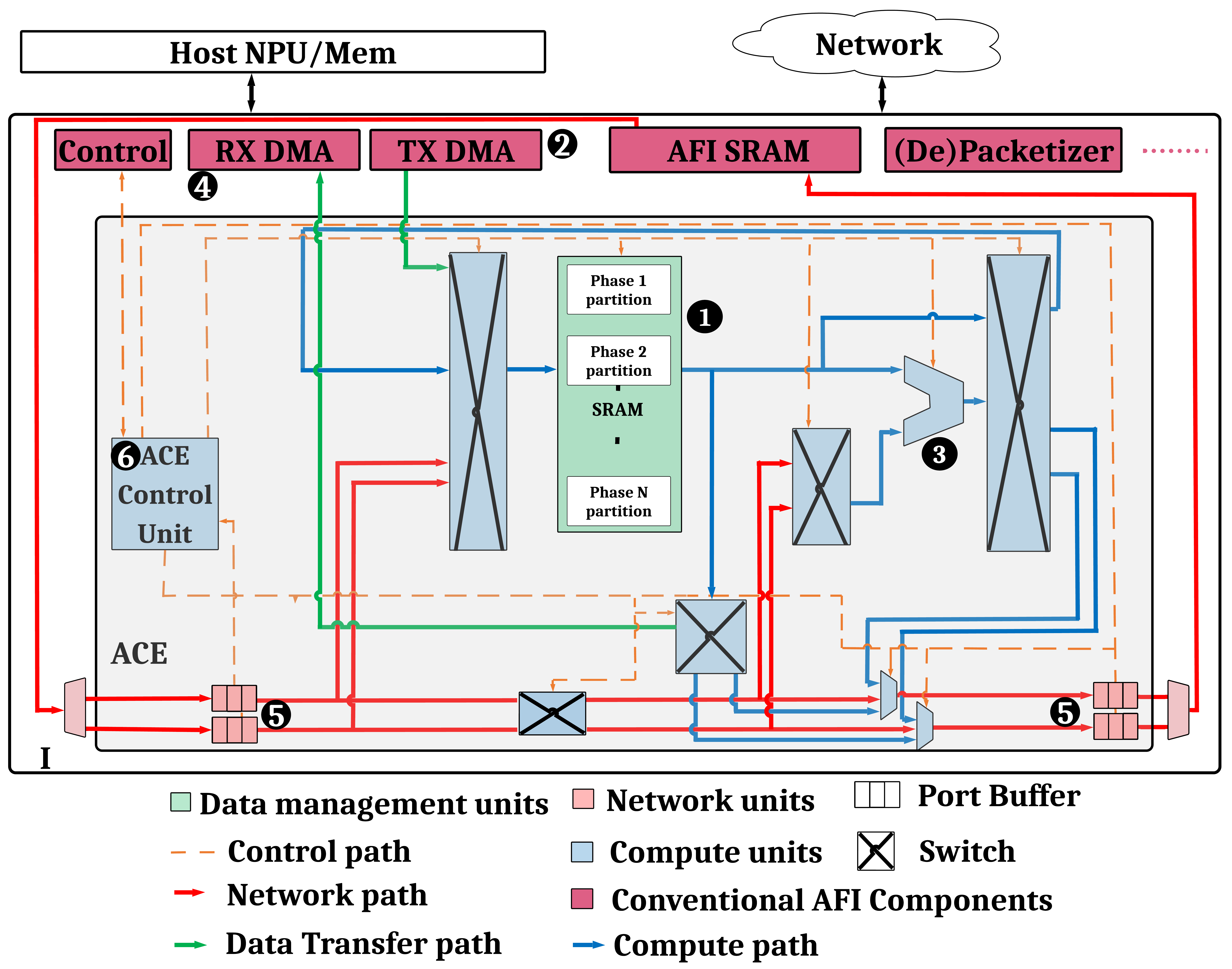}{\ncu microarchitecture. \#1 is the on-chip SRAM. \#2 is the AFI TX DMA for transferring data from main memory to AFI SRAM (normal operation) or \ncu SRAM (\ncu activated). \#3 is the ALU. \#4 is the AFI RX DMA for transferring data from AFI SRAM (normal operation mode) or \ncu SRAM (\ncu activated mode) to main memory. \#5 are the input/output port buffers.  These buffers are allocated per each physical link and contain packets corresponding for that specific link. \#6  is the control unit logic.}{0.8}

\begin{table}[t]
\caption{Data granularity at different levels of \ncu execution.}
\resizebox{1.0\linewidth}{!}{\begin{tabular}{|l|l|l|}
\hline
\multicolumn{1}{|c|}{\textbf{Granularity}} & \multicolumn{1}{c|}{\textbf{Size}}                                              & \multicolumn{1}{c|}{\textbf{Constraint}}                                    \\ \hline
Payload (variable)                                   & Training Algorithm                                                              & Training Algorithm                                                          \\ \hline
Chunk (64kB initially)                                     & Parameter for Pipelining                                                        & \begin{tabular}[c]{@{}l@{}}Storage Element Size\\ (Area/Power)\end{tabular} \\ \hline
Message (4kB)                                  & \begin{tabular}[c]{@{}l@{}}Parameter - \\ Multiple of Number of Nodes\end{tabular} & Topology                                                                    \\ \hline
Packet  (256B)                                   & Link Technology                                                                 & Technology                                                                  \\ \hline
Flit  (256B)                                     & Network Buffer Size                                                                     & \begin{tabular}[c]{@{}l@{}}Microarchitecture\\ (Area/Power)\end{tabular}    \\ \hline
Phit (variable)                                      & Link Width                                                                      & Technology                                                                  \\ \hline
\end{tabular}}
\label{table:data_granularity}
\end{table}

\section{Accelerator Collectives Engine}
\label{sec:design}
\insertWideFigureNew{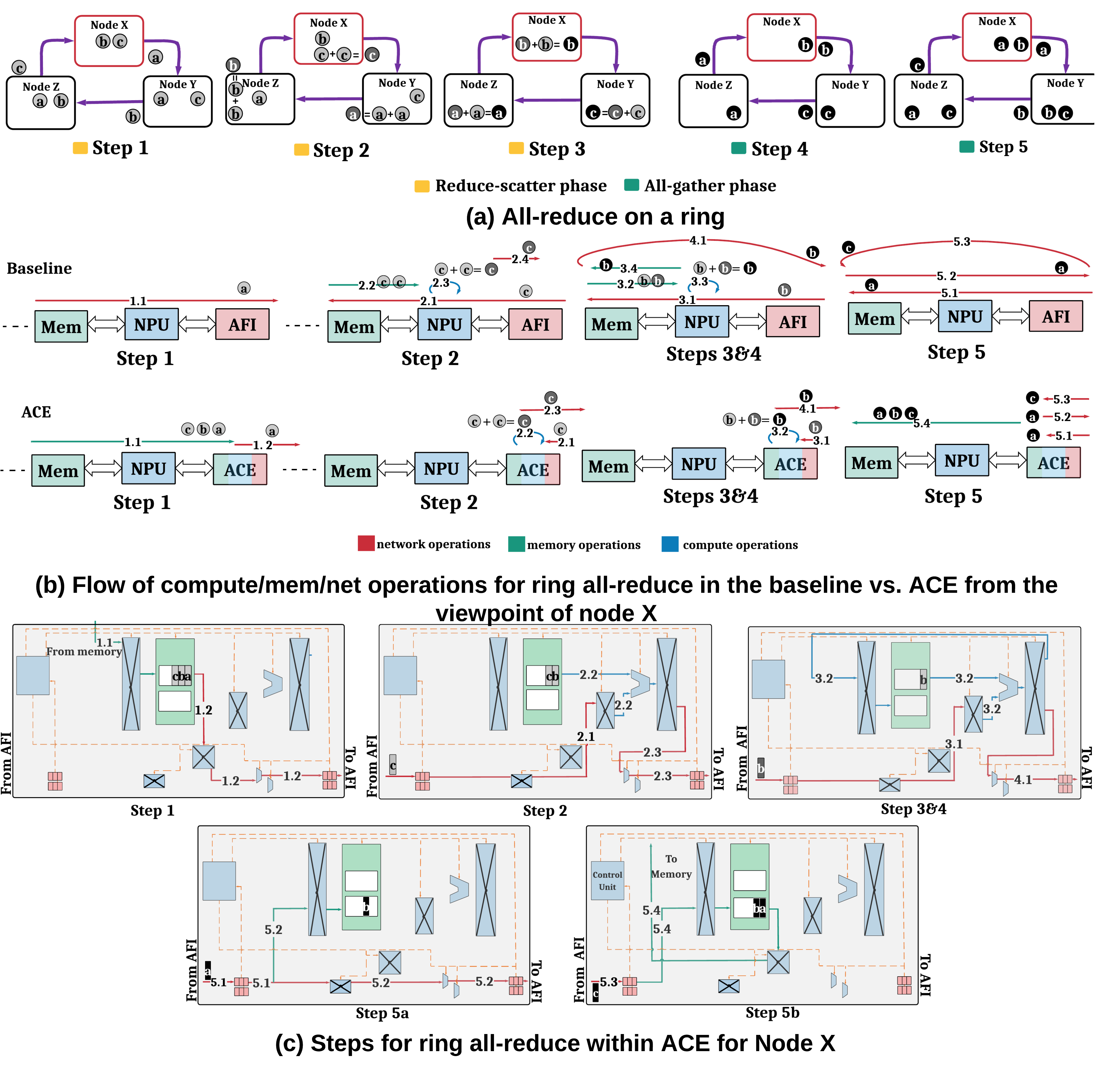}{Implementation of ring all-reduce on Baseline vs ACE}{0.85}

Driven by the 
compute and memory 
contention issues identified by \autoref{sec:motivation}, 
this work makes a case for 
a dedicated accelerator for DL collective communication called \ncu, sitting alongside the Accelerator Fabric Interface (AFI) module introduced earlier in \autoref{fig:overviewOfServer}. 
We present design details next.

\vspace{-2mm}
\subsection{System Overview}
\autoref{fig:microarch} shows the high-level overview of \ncu integrated into the AFI module. 
The key components within the \ncu are additional compute units (for all-reduce or all-to-all computations)
and scratchpad buffers (to cache gradients).
Like other domain-specific accelerators~\cite{TPU},
the datapath of \ncu is 
specialized for its target workload - namely running collective communication algorithms.

\subsection{Placement: End-point vs Switch}
\label{sec:endpoint_vs_switch}
The communication offload solutions are not new.
Fundamentally, there are two options for offload:
(i) switch-based, and (ii) endpoint-based offload methods. 
\autoref{tab:SwitchVsEndpoint} shows the comparison between these two. Both methods reduce the burden on endpoint memory and compute for handling communication-related tasks. However, endpoint-based methods are more flexible in supporting various network topologies, as opposite to switch-based methods that: (i) require a switch in the network topology, and (ii) point-to-point (PTP) links do not benefit from the offload scheme. Note that in AFs, many platforms are designed solely using point-to-point network topologies (e.g. NVIDIA DGX-1 \cite{dgx1}, Google TPU \cite{TPU}, Habana \cite{HabanaPtP}. This is in addition to the intra-package networks that are almost always PTP  \cite{SiliconInterposer}. Additionally, unlike switch-based methods that only support a single reduction/broadcast traffic pattern to/from the switches, endpoint-based methods can support many different algorithms (patterns) (e.g. ring, halve-doubling, double-binary-tree, etc.). The above reasons show why we employ \ncu as an endpoint offload, sitting alongside the AFI.



\vspace{-1mm}
\subsection{Data Granularity}
\label{sec:data_granularity}
\autoref{table:data_granularity} shows the granularity of data at different levels of the \ncu execution and their determining factor. It also shows the default value of each level used in \ncu. \ncu initiates execution by receiving a command from NPU to perform a specific collective on a \emph{payload}. The payload could be 
activations or gradients depending on the parallelism approach and forward/back pass. 
The command includes the collective type and the address range for data residing in the main memory. \ncu then divides the payload into multiple \textit{chunks} and begins processing and scheduling of each chunk individually and in a pipelined manner.
Multiple chunks can be scheduled in parallel.

A chunk itself decomposes into multiple \textit{messages} and the collective algorithm runs at message granularity.
The number of messages is a multiple of the number of nodes in the system.
For example, if \ncu wants to perform an all-reduce in a ring with 4 NPUs, it can divide the chunk into 8 messages, 
and execute all-reduce serially over two iterations\footnote{Each step leads to processing \& performing all-reduce for a group of 4 messages and the algorithm is ring-based. More details on ring-based all-reduce is provided in \autoref{sec:walkthrough}.}.


Each message comprises one or more \textit{packets} when it enters the network layer. 
The unit of data processing within the \ncu is packets. The bus width and SRAM interface might or might not be equal to the size of the packets and data movement/execution is serialized if the size is smaller than packet width.

\subsection{Walk-Through Example for All-Reduce}
\label{sec:walkthrough}

We describe \ncu in action via a detailed walk-through example for running the ring-based all-reduce collective over a ring for both  baseline and \ncu as shown in \autoref{fig:walkthrough_combined}.
The general concepts and the main advantages of \ncu compared to the baseline are applicable to any topology/collective, as we describe later in \autoref{sec:flexibility}.

\autoref{fig:walkthrough_combined}.a shows the logical flow of the algorithm across the different nodes. We assume one chunk for simplicity.
Since there are three nodes, there are three messages\footnote{Note that there could be multiple number of 3 messages and as we described in \autoref{sec:data_granularity}, they should be executed in serial. But here for simplicity we assume only 1 group of 3 messages.}.
An all-reduce can be implemented as a reduce-scatter followed by an all-gather, as can be seen from \autoref{fig:collectives_table}. 
Steps 1-3 show the reduce-scatter part. Step 1 initiates the reduce-scatter; each node  sends one message to its neighbor and waits for receiving another message from its other neighbor. In step 2, each node reduces the received message with its local one and forwards it to the next. Step 3 concludes reduce-scatter by  each node reducing the last message it has received. All-gather starts with each node forwarding a copy of its reduced message to its neighbor (step 4) and then each node keeping a copy of its received message and forwarding it (steps 5, 6).

\autoref{fig:walkthrough_combined}.b shows this flow from node X's view in the case of baseline vs. \ncu. It is clear from this figure that in baseline, in all phases, messages need to go all the way from/to main memory to/from AFI to be injected/received into/from the network. The local reduction in baseline begins by loading the local message and the received message into the NPU cache (e.g. sub-step 2.2),  performing reduction (e.g. sub-step 2.3), and sending the result out to the network (e.g. sub-step 2.4) or main memory (e.g. sub-step 3.4), depending on the step of the algorithm. This in turn reduces the available memory bandwidth and compute resources for the 
DL GEMMs.
In contrast, \ncu restricts the data movement only to the first and last phases (reduced congestion and increased available memory bandwidth) and allows the DL GEMMs 
to make use of complete NPU compute resources.  

\autoref{fig:walkthrough_combined}.c shows the internal \ncu interactions for node X. Here, the \ncu SRAM is divided into two partitions - one serves as a source for the (only one) all-reduce phase, and the last one serves as the source to hold the final results to send back to main memory.
%
In step 1, the 3 messages are brought into the first partition of \ncu SRAM by the TX DMA (sub-step 1.1). Then, one message is sent out through a series of packets\footnote{Note that here packets means packet data. The actual packetization of this data is the job of AFI once it wants to send it over the links.} injected into the designated output port buffer to be injected  into the network (sub-step 1.2). In step 2, the received message is reduced  with the local data (sub-step 2.2) and forwarded to the neighbor (sub-step 2.3). \ncu overlaps steps 3 and 4 of the algorithm; after performing a reduction (sub-step 3.2), stores it locally (sub-step 3.2) and forwards it to the next neighbor (sub-step 4.1) at the same time. Step 5 is broken into two figures for more clarity. In step 5a, the received message is stored and forwarded (sub-step 5.2) at the same time, while in step 5b, the final received message is stored and the whole chunk is sent back to memory by RX DMA (sub-step 5.4).

It is clear that in some steps within \ncu, multiple resources should be available for some sub-steps to proceed. For example, in sub-step 5.2 in step 5a, both the SRAM input port should be available and the output port should have free space. In such cases, if any of the resources are not free, that step is stalled until all resources are available. Multiple chunks can be executed in parallel to maximize internal hardware resources and link bandwidth, as we discuss next.

\subsection{Parallelism}
\label{sec:parallelism}
To achieve high network utilization, we need to apply parallelism at various levels. From the algorithmic perspective, there are several levels where parallelism is possible. 
More complex hierarchical topologies 
implement topology-aware collectives over multiple phases~\cite{astrasim}.
Assuming the collective algorithm has P phases, multiple chunks can run in parallel both within a phase and across different phases. Each chunk will send/receive multiple messages. Hence, multiple in-flight chunks mean multiple in-flight messages (belonging to different chunks) are possible. Parallel chunks mean multiple packets can be processed in parallel. Packets are the unit of data transfer within the network and parallelism below that is the network's job. So the \ncu memory management and control units are designed to ensure using all algorithmic parallelism opportunities.
%
The SRAM within \ncu is partitioned according to the number of phases of the collective algorithm being run plus one for holding the final results for RX DMA, called the \emph{terminal partition}.
For the example in \autoref{sec:walkthrough}, the ring-based all-reduce has one phase and hence needs two partitions. More details on multi-phase all-reduce are mentioned in \autoref{sec:methodology}.
\subsection{Control Unit}
\label{sec:control}

The control unit comprises multiple programmable finite state machines (FSMs).
Each FSM can be programmed for a specific phase of a specific collective algorithm and holds a queue of chunks that should process in order. Each entry of this queue holds the \emph{context} of a chunk like its start and end address inside the SRAM and the address range for holding the final result for the next phase. When a chunk is created in \ncu, it is also assigned the state machines it should go through for each phase.\footnote{It is possible that, for a given workload and parallelism (e.g., DLRM in our evaluations), different collective operations (e.g. all-reduce and all-to-all) exist for the same phase. In this case, the FSMs allocated for that phase should be programmed to handle all collective operations for that phase.}.
%
%
When entering each phase, the chunk is inserted into the queue of its assigned state machine for that phase. The state machines then compete with each other to access different resources, resulting in overlapping and out-of-order execution of multiple chunks both within\footnote{If the phase is assigned multiple FSMs.} and across phases. This increases resource utilization and network bandwidth. The available parallelism is only bounded by the number of available state machines to manage the dataflow for each phase.

\subsection{Interface with NPU and AFI}
\ncu extends the existing AFI interface exposed to NPU as shown in \autoref{fig:microarch}.
AFI control forwards the \ncu-specific commands from NPU/\ncu to \ncu/NPU. The NPU-AFI command interface is similar to UCX \cite{UCX} or OFI \cite{OFI} which are the standard high-level communication interfaces. Once a collective command is received, \ncu decides when to load data given the availability of SRAM space. Finally, \ncu notifies the completion of chunk collective by raising an interrupt and forwarding it to NPU.
%


\vspace{-2mm}
\subsection{Flexibility and Scalability}
\label{sec:flexibility}

\textbf{Supporting Different Collectives.}
The general principles for running any collective algorithm using \ncu remain the same. For a collective with say P phases, the SRAM is divided into P+1 partitions. Each partition is assigned to a phase and one or multiple FSMs, except the last partition (i.e. terminal partition) that is used for storing the final results to be written back to memory.
Different collectives can be implemented by programming the dataflow into the FSM.

\textbf{Supporting Different Topologies.}
Sine \ncu handles the collectives at the endpoints, it is orthogonal to any network topology (e.g., switch-based, point-to-point, hybrid). From the logical view, \ncu can perform any collective algorithm (e.g. ring-based all-reduce) on top of any physical topology. It is then the job of network protocol and routing algorithm to deliver the packets accordingly.

\begin{table}[h]
\caption{Synthesis Results}
\vspace{-2mm}
\centering
\resizebox{0.8\linewidth}{!}{\begin{tabular}{|l|l|l|}
\hline
\multicolumn{1}{|c|}{\textbf{Component}} & \multicolumn{1}{|c|}{\textbf{Area} ($\mu m^2$)} & \multicolumn{1}{|c|}{\textbf{Power} (mW)} \\ \hline
ALU & 16112 & 7.552 \\ \hline
Control unit & 159803 & 128 \\ \hline
4$\times$1MB SRAM banks & 5113696 & 4096 \\ \hline
Switch \& Interconnect & 1084 & 0.329 \\ \hline
ACE (Total) & 5339031 & 4255 \\ \hline
\end{tabular}
\label{tab:synth}
}
\vspace{-2mm}
\end{table}

\subsection{\ncu Design-space Exploration and Implementation} 

\autoref{fig:ACE_DSE}.a shows the \ncu performance when SRAM size and number of FSMs, the components with the most overheads, are swept. \autoref{fig:ACE_DSE}.a shows that increasing beyond 4MB of SRAM and 16 FSMs results in diminishing returns, and so we chose these as our final parameters\footnote{Only 6$\%$ performance improvement is seen for 8MB SRAM and 20 FSMs.} since the selected parameters are enough to fill most of the network pipeline.
Also, 4 wide ALU units, each capable of preforming 16$\times$FP32 or 32$\times$FP16 in parallel were sufficient for \ncu. The interconnect between SRAM and functional units are wide 64B buses. We implemented ACE using Verilog and synthesized the design using the Synopsis Design Compiler in 28nm technology node. \autoref{tab:synth} shows the area and power estimates for our design, enumerating individual components as well as ACE itself. Compared to the area and power of high-end training accelerators reported in \cite{TPU,AcceleratorBenchmark}, \ncu has less than $2\%$ overhead in both area and power.

\insertFigureNew{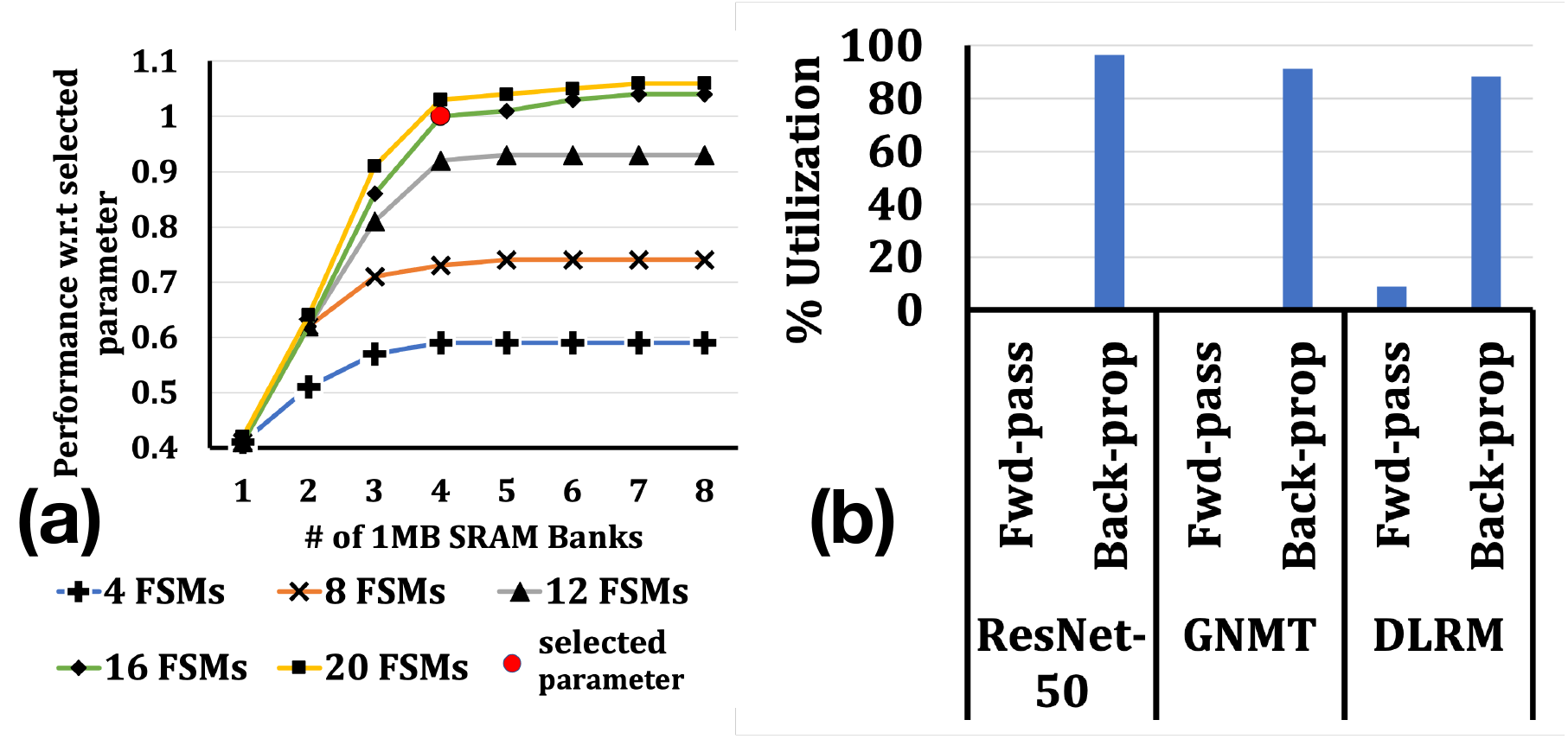}{(a) The performance of ACE with different FSM and SRAM sizes normalized to the performance of selected parameters (i.e. 4MB SRAM, 16 FSMs). The performance is gathered through averaging ACE performance across all target workloads and system sizes described in \autoref{sec:methodology}. (b) The utilization of \ncu for the simulations of \autoref{fig:detailed}. \ncu is considered utilized when it has assigned at least one chunk for processing.}{0.9}

In addition, we used a simple heuristic for SRAM partitioning that partitions the SRAM based on the (available network bandwidth $\times$ initial chunk size) for each phase\footnote{For example, a phase with 2$\times$ link bandwidth and 2$\times$ initial chunk size has a partition 4$\times$ greater than a phase with 1$\times$ bandwidth and 1$\times$ chunk size. Also note that if for each phase, there are different chunk sizes belonging to different collective operations, we use average of such chunk sizes.}, with the \emph{terminal partition} (i.e. partition P+1) being equal to the the last phase partition (i.e. partition P).

\begin{table}[htbp]
\caption{System parameters}
\vspace{-3mm}
\label{tab:SystemParameters}
\begin{center}
\footnotesize
\begin{tabular}{|c|c|}
\hline
\textbf{Parameter} & \textbf{Values} \\
\hline
Compute Accel. & \makecell{Max of 120 TFLOPs FP16 at\\ GPU-like, with 80 SMs}\\
\hline
Bus BW & 500 GB/s (NPU-AFI), 900 GB/s (NPU-MEM) \\
\hline
\ncu Message size & 8KB \\
\hline
Packet size & 256 Bytes (Intra \& Inter Package) \\
\hline
Per link BW & 200 GB/s (Intra), 25 GB/s (Inter) \\
\hline
Link latency & 90 cycles (Intra), 500 cycles (Inter) \\
\hline
Total \# of links/NPU &  \makecell{2 intra-package links (1 bidirectional ring),\\ 4 inter-package (1 bidirectional horizontal ring,\\ 1 bidirectional vertical ring)} \\ 
\hline
Total BW &  \makecell{400 GB/s (intra-package ring),\\ 50 GB/s (horizontal ring),\\50 GB/s (vertical ring)} \\ 
\hline
Link efficiency & 94\% (Intra \& Inter Package) \\   
\hline

\hline
\end{tabular}
\label{tab:system_parameters}
\end{center}
\end{table}

\begin{table*}[h]
\caption{Target System Configurations}
\footnotesize
\centering
\begin{tabular}{|p{8em}|p{55em}|}
\hline
\textbf{BaselineNoOverlap} & State-of-the-art baseline system where there is no overlap between communication and computation. All communications are gathered and explicitly issued at the end of back-propagation phase using a single communication kernel*. This means that all resources are available for compute and communication when they are running. This strategy enhances compute time but puts all communication time on the critical path since training loop can not proceed before the communication is finished. \\ \hline
\textbf{BaselineCommOpt} & State-of-the-art baseline system where enough resources are allocated to make its communication performance reaches 90 \% of Ideal case. Using the intuition behind \autoref{fig:motiv} and \autoref{fig:motiv2}, 450 GB/s of memory BW and 6 SMs of the NPU is allocated for the communication task and the rest for the computation. \\ \hline
\textbf{BaselineCompOpt} & State-of-the-art baseline system, but this time training computation performance is optimized. In order to do this and similar to \ncu, we assume only 128 GB/s of memory BW is allocated for communication. Since this amount of memory BW is not enough to derive the entrie BW (see figure \autoref{fig:motiv}), then using results of \autoref{fig:motiv2}, 2 SMs are enough for the communication task. The rest is for compute. \\ \hline
\textbf{\ncu} & Proposed system. In this case, 100\% of the NPU resources are dedicated to the training algorithm computation. Moreover, 128 GB/s memory BW is enough to reach to 90\% of the ideal (see \autoref{fig:motiv}). The rest of memory BW is allocated for training compute.  \\ \hline
\textbf{Ideal} & A system where the endpoint can handle/ process received messages magically within one cycle. This essentially means that there is no associated latency from the endpoint side in the collective communication latency. This gives an upper bound to our design. In this case 100\% of compute and memory is allocated for training algorithm only. \\ \hline
\end{tabular}
\label{tab:target_systems}
*The only exception is DLRM fwd-pass all-to-all where the training loop performs a blocking wait until the communication completes and then starts compute.
\end{table*}

\section{Evaluation Methodology}\label{sec:methodology}
This section describes our methodology establishing and simulating  high-performance training systems and evaluating the benefits of communication acceleration.

\textbf{Simulator.}
\autoref{tab:system_parameters} shows the major system parameters.
We used ASTRA-SIM \cite{AstraSimGithub,astrasim}, a distributed DNN training simulator 
and developed \ncu on top of that.
ASTRA-SIM models the training loop,
parallelization, and
collective communication scheduling similar to libraries like oneCCL~\cite{oneCCL}.
It interfaces with a compute model~\cite{Scalesim} for the compute times (i.e. forward-pass, weight gradient, and input gradient) and network simulator~\cite{Garnet} for tracking cycle-level network behavior.

\textbf{Compute Parameters.}
%
The compute model (NPU) is a 1245 MHz GPU-like core with 80 Streaming multiprocessors (SMs) that can achieve the max of 120 TFLOPs using FP16 precision. Our compute accelerators is quite similar to the state-of-the-art training accelerators \cite{Volta}.

\textbf{AFI and Network Parameters.}
We extended ASTRA-SIM to model AFI and the interactions (traffic) between AFI, NPU, and Memory. We also modeled the transaction scheduling effects of NPU-AFI and NPU-Mem and queuing delays of subsequent transactions. NPU-Mem bandwidth is chosen based on the state-of-the-art accelerators \cite{Volta}. NPU-AFI bandwidth is assumed to be the sum of all intra-package/inter-package links as a logical choice to prevent NPU-AFI bandwidth to be the bottleneck in driving the network links. Inter-package link BW is assumed to be the same as NVlink \cite{NVlink}, while intra-package bandwidth is selected based on \cite{Simba,IntelGPU} that is an expected bandwidth for high-performance MCM systems.




\textbf{Target Training Platforms.}
\ncu works for both pt-to-pt and switch-based topologies. In the interest of space, we present the results for a pt-to-pt
3D Torus topology.
We model the futuristic platforms with AF comprising of multiple NPUs integrated through multi-chip packaging technology \cite{IntelGPU,DLCircuit} on a package, and multiple packages interconnected via a dedicated fabric. We use the notation LxVxH to describe the system size, where L stands for the number of NPUs within the same package that are connected through an intra-package ring (local dimension). All NPUs with the same offset across different packages then form a separate 2D torus within V rows (vertical dimension) and H columns (horizontal dimension), connected through vertical and horizontal rings, respectively.



\textbf{Topology-aware Collective Algorithms.}
We use hierarchical multi-phase collective algorithms for the 3D torus topology, where the \textit{\textbf{all-reduce}} occurs in 4 phases: reduce-scatter in local, all-reduce in vertical, all-reduce in horizontal followed by all-gather in local dimension. This implementation uses the higher-bandwidth intra-package local links more than the inter-package links and is provided by the simulator~\cite{astrasim}. For the \textit{\textbf{all-to-all}} collective, we used the direct all-to-all where each NPU simultaneously sends a distinct portion of data to any other NPU \cite{TPU,hotiPaper}. Since all-to-all is a single-phase, all FSMs are programmed to be able to execute all-to-all, in addition to their assigned all-reduce phase. To perform all-to-all, each FSM places the data to the output link FIFOs based on the destination NPU route. We used XYZ (local dim, vertical dim, horizontal dim) routing for each packet to reach its destination for our torus-based topologies. 
Note that AF networks like NVLink only support
neighbor-to-neighbor communication natively. This means that for the baseline system and for the packets that need to go multiple hops, the communication library (e.g. NCCL) is responsible for writing data to the intermediate hops’ memory and again forwarding data to the next hop. This wastes a lot of memory BW on the intermediate hops. But ACE prevents such unnecessary memory overheads since its SRAM absorbs packets and forwards the ones that have different destinations through the FSM responsible for the corresponding chunk.



\textbf{Target System Configurations.}
We consider five different system configurations, discussed in \autoref{tab:target_systems}.
We investigate three flavors of the baseline systems: two of them with comp/comm overlap either optimized for communication (BaselineCommOpt) or computation (BaselineCommOpt), and one other baseline where there is no overlap between communication and compute (BaselineNoOverlap). In addition to \ncu, we also present the Ideal system results to get the upper bound if we had maximum compute/communication performance.

\textbf{Target Workloads.}
In order to evaluate our platforms, 
we consider three different sets of real workloads: (1) ResNet-50 \cite{Resnet} from vision, GNMT \cite{gnmt} from NLP, and DLRM \cite{dlrm} from recommendation DNNs. We use the FP16 precision for activation/gradient computation and communication. We consider data-parallel parallelism (i.e. requiring all-reduce for weight gradients) for ResNet-50 and GNMT and hybrid parallel (data-parallel across MLP layers, model parallel across embedding tables) for DLRM. For DLRM, we use the version described in \cite{hotiPaper} which is a representative model used in real systems. We simulate two training iterations with Last-In-First-Out (LIFO) collective scheduling policy to give more priority to the collectives of first layers during back-propagation. The mini-batch size is set to be 32, 128, and 512 per NPU for ResNet-50, GNMT, and DLRM, respectively. We present both end-to-end runtime and 
breakdown of compute-comms overlap. We assume weak scaling as we scale the number of NPUs in our evaluations.

\textbf{Metric of Evaluation.} For real workloads (\autoref{sec:compCommOverlap}, \autoref{sec:scalability}, and \autoref{sec:DLRMOptimization}), our metrics are \emph{total computation} and \emph{exposed communication}. Exposed communication refers to the time where communication cannot be overlapped with computation and hence, the training algorithm (compute) is forced to stop because it is 
waiting for the communication to be finished. This case for data-parallel is during forward pass, where for each layer we need to make sure the weight gradient communication (i.e. all-reduce) of the previous iteration is completed and weights are updated. For DLRM, we additionally need to wait for the embedding communication operation (i.e. all-to-all) before entering the top MLP layers in forward pass, and after finishing of the back-propagation to update the embedding \cite{dlrm}. The summation of the (total computation + exposed communication) then determines the training iteration time.
\vspace{-1mm}
\section{Evaluation Results}\label{sec:results}
\vspace{-1mm}

This section presents the simulation results comparing ACE against \emph{Ideal} and the baseline systems mentioned in \autoref{tab:target_systems} for real workloads. But first, we do an analytical investigation about the memory BW requirement for \ncu vs. Baseline, that justifies the simulation results we observed in \autoref{fig:motiv}.


\subsection{Memory BW Requirements for Baseline vs. \ncu }\label{sec:analyticalBound}
According to \autoref{fig:walkthrough_combined}, in the baseline system, the number of memory reads is 2N bytes for sending every N bytes to the network in the reduce-scatter phase (excluding the initial phase). N memory bytes need to be read to send N bytes to the network for all-gather phase. Since the number of bytes to be sent out in these 2 parts is identical, this means that on average, 1.5N bytes need to be read from memory to send out N bytes. This explains there is a need for $\approx$450 GB/s memory BW to drive 300GB/s of the network, in the ideal case. 

However, \ncu caches the data and reuses it multiple times. The amount of reuse depends on the topology and algorithm. Let's consider the 64-NPU (4X4X4) system, similar to \autoref{fig:motiv}. For every N bytes of data that is cached in \ncu, $\frac{3}{4}$N is sent out during the first phase of reduce-scatter, 2$\times\frac{6}{16}$N is sent out one for vertical and one for horizontal all-reduce phases. Finally, another $\frac{3}{4}$N is sent out for the final all-gather phase, resulting in a total of 2.25N data sent to the network (133 GB/s memory BW to drive 300 GB/s in the ideal case). However, in reality, more memory BW is needed due to hardware delays, queuing delays on the buses, etc., as we showed in \autoref{fig:motiv}. We limit our analytical study to all-reduce collective due to the limitation of space, but a similar analysis can be done for all-to-all as well. However, compared to the all-reduce, all-to-all is used much less frequently and their sizes are usually smaller.

\insertWideFigureNew{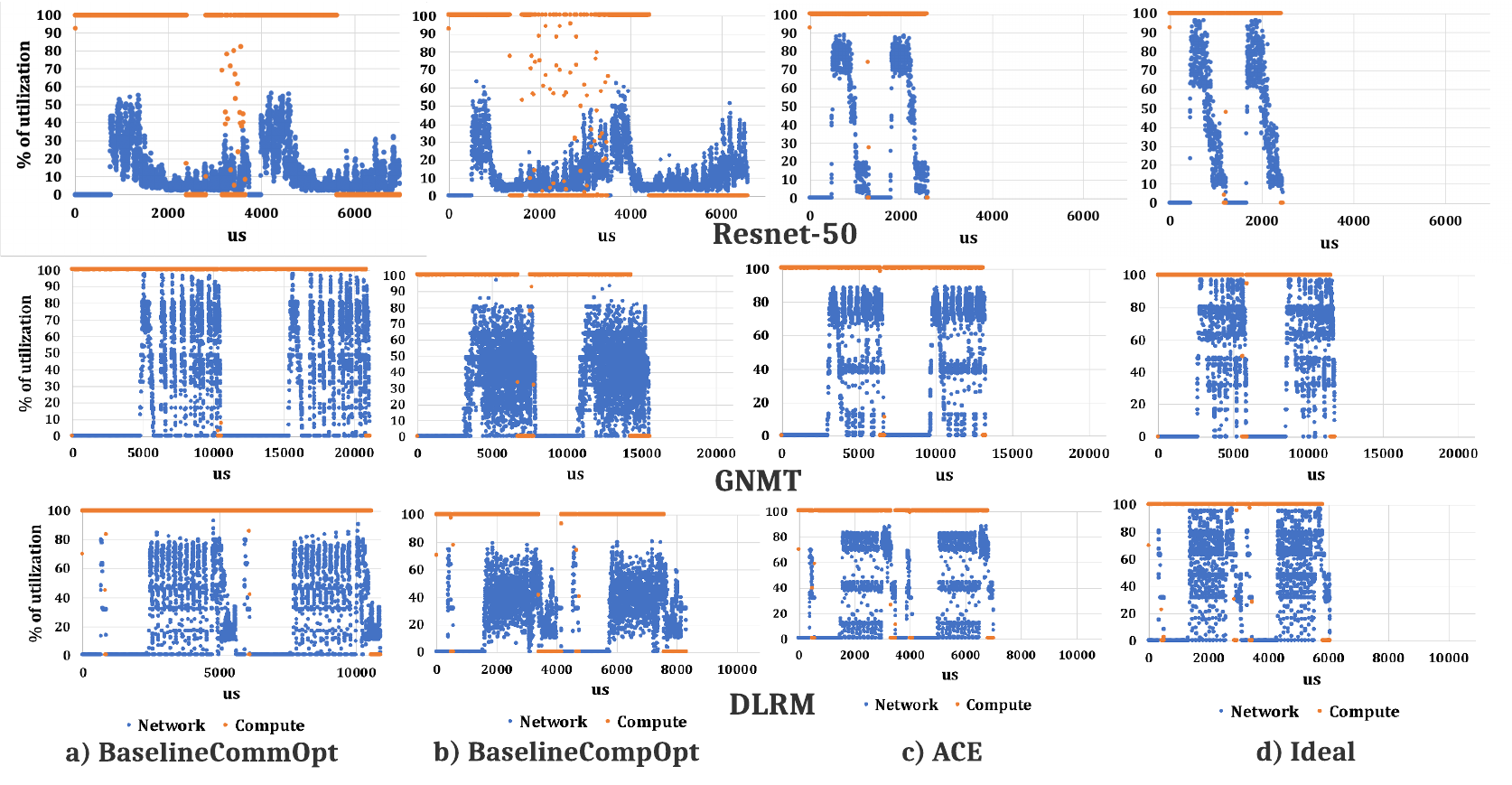}{Compute-Communication overlap for 2 training iterations of ResNet-50,GNMT and DLRM networks running on a 4x8x4-node 3D Torus. Each point reports the average compute/network utilization over 1K cycles (500 cycles before, and 500 cycles after) period of training. There are 2 bursts of network activity corresponding to 2 iterations. \textbf{Note that here, \% of network utilization corresponds to the \% of links that utilized for scheduling a flit in a cycle, irrespective of their different link BW.}}{0.97}

\insertWideFigureNew{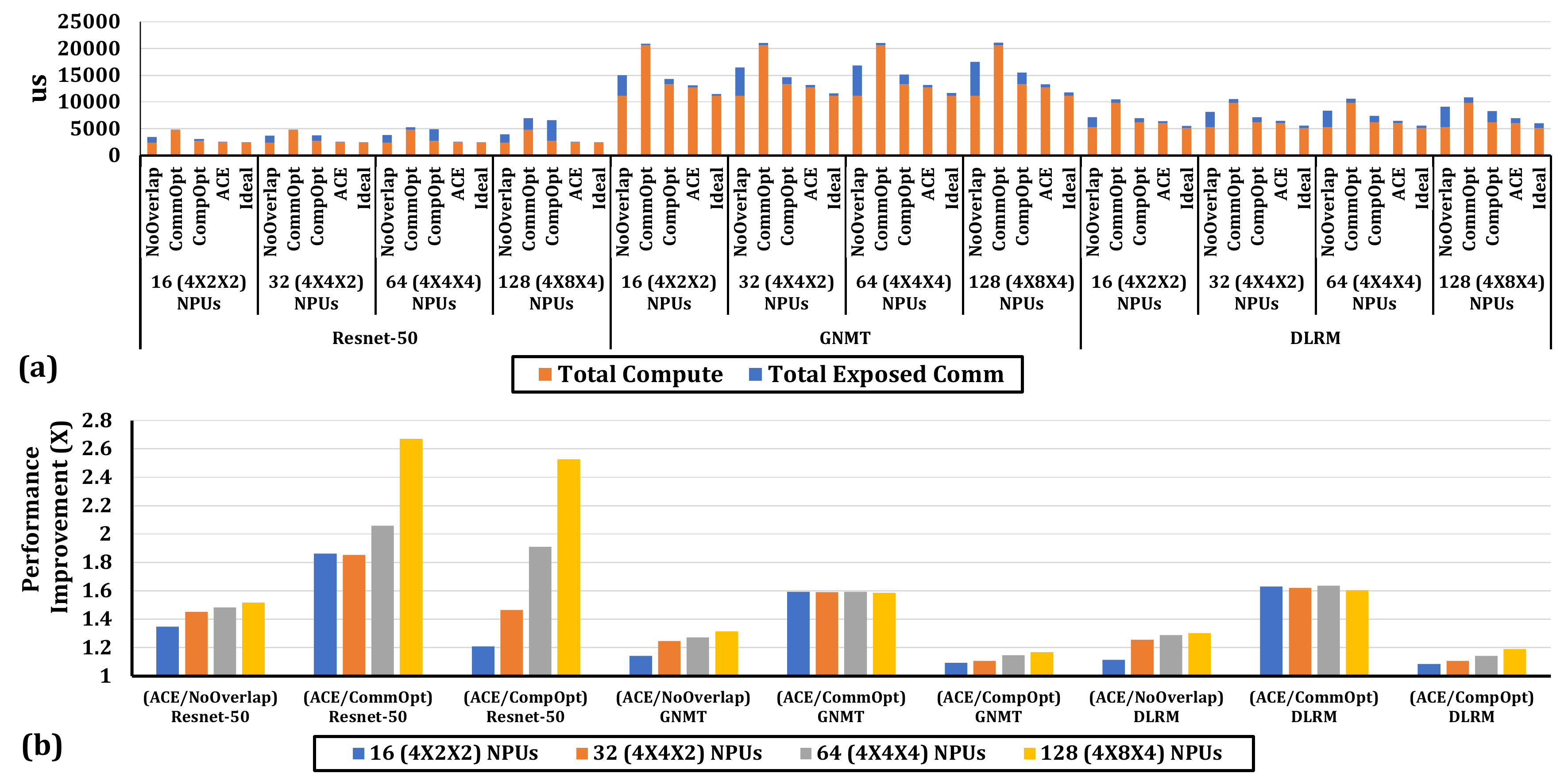}{a) Total computation time vs. total exposed communication (for 2 training iterations) time as the number of NPUs in the AF network increases. b) The corresponding performance of \ncu over BaselineNoOverlap, BaselineCommOpt and BaselineCompOpt.}{0.90}


\subsection{Compute-Communication Overlap}\vspace{-1mm}\label{sec:compCommOverlap}

Here, we evaluate the four systems with compute-communication overlap mentioned in \autoref{tab:target_systems} for \textbf{two} training iterations on ResNet-50, GNMT, and DLRM. The goal is to dissect the compute-communication overlap of different configurations for a single 128 NPUs AF network size. In \autoref{sec:scalability}, we show the general trend for different network sizes and include BaselineNoOveralap configuration as well.

\textbf{ResNet-50 Analysis.}
The first row of \autoref{fig:detailed} shows the NPU compute and network utilization for ResNet-50 training.

\textbf{\textit{BaselineCommOpt.}}
First, we focus on the BaselineCommOpt (\autoref{fig:detailed}.a). Here, we observe that: i) despite being optimized for communication, network links are still not yet completely utilized, and ii) network utilization also fluctuates at different times.
This is because of three main reasons. First, in real workloads there are some computation operations between two consecutive communication, allowing for partially/completely finishing current communication before the next communication task arrives. Second, high BW of the intra-package links makes them underutilized in the collective algorithm. Additionally, Resnet-50 issues many small-size collectives that make it not sufficient to completely drive the available network BW\footnote{Note that because of these reasons, even in the Ideal case the network utilization fluctuates.}. Thus, exposed communication still consists of 31.5$\%$ of iteration time.

\textbf{\textit{BaselineCompOpt.}} 
Next, we dive into the second flavor of the baseline that is optimized for compute. Compared to BaselineCommOpt, BaselineCompOpt reduces the total computation time by 1.75$\times$ but it fails to significantly reduce iteration time since exposed communication increases due to the poor utilization of the network. Note that in \autoref{fig:detailed} the network utilization of BaselineCommOpt and BaselineCompOpt seems similar. This is because due to the much shorter intervals between the communication tasks in BaselineCompOpt, as a result of the reduced computation, that allows it to schedule larger size collectives at a given time. The overall iteration time is improved only by 1.06$\times$, compared to the BaselineCommOpt.

We note that this result does not mean that our baseline system has poor performance.  
In fact, as \autoref{fig:detailed}.b shows (and we summarize later in \autoref{fig:scale_results})
one training iteration duration takes $\sim$ 3.5 ms for the 128-node baseline system. This means that one epoch of training on the ImageNet dataset \cite{imagenet} takes $\sim 12$ seconds that is comparable to the best training performance-per-node times reported \cite{FastTraining}. However, we will show that releasing more resources for compute via \ncu leads to even better performance.

\textbf{\textit{\ncu.}}
\ncu can achieve the best of both worlds by releasing most of the resources for compute and still being efficient through caching the data and performing the collective. Compared to the BaselineCompOpt, it can slightly improve compute time by 1.06$\times$ because of the releasing all compute cores for compute. However, it still can handle the communication efficiently and exposed communication is only $2\%$ of the iteration time. \ncu outperforms BaselineCommOpt and BaselineCommOpt by 2.67$\times$ and 2.52$\times$ in terms of iteration time, respectively. Overall, \ncu can achieve 94.3$\%$ of ideal system performance, while this value is 35.3$\%$ and 37.3$\%$ for BaselineCommOpt and BaselineCompOpt, respectively.

\textbf{GNMT Analysis.}
The second row of \autoref{fig:detailed} shows the compute and communication utilization for GNMT. In GNMT, communication sizes (per layer) are larger allowing BaselineCommOpt to drive the network efficiently. Also, larger compute time means more room to overlap communication with computation. However, BaselineCommOpt still suffers from the long compute time especially since GNMT compute is more sensitive to available memory BW. \ncu can outperform both BaselineCommOpt and BaselineCompOpt by 1.59$\times$ and 1.17$\times$, respectively. Additionally, it achieves 88.7$\%$ of the ideal system, compared to the BaselineCommOpt and BaselineCompOpt with 55.9$\%$ and 76$\%$.

\textbf{DLRM Analysis.}
The third row of \autoref{fig:detailed} corresponds to the DLRM results. Again, here the communication size is relatively large. \ncu iteration time is shorter by 1.55$\times$ and 1.19$\times$ compared to BaselineCommOpt and BaselineCompOpt, respectively. BaselineCommOpt,  BaselineCompOpt, and \ncu achieves 55.6$\%$, 72.8$\%$, and 86.5$\%$ when compared to the ideal system iteration time, respectively.

\textbf{\ncu Utilization.}
\autoref{fig:ACE_DSE}.b shows the average utilization of \ncu. As can be seen in \autoref{fig:ACE_DSE}.b, for forward-pass,  ResNet-50 and GNMT have zero communication while DLRM has only a single all-to-all communication, resulting in low utilization of \ncu. However, for back-propagation, the average utilization for ResNet-50, GNMT, and DLRM is 96.4$\%$, 91.3$\%$, and 88.3$\%$, respectively. The reason for not full utilization is the presence of compute between communication task and high-performance of \ncu that results in some idle times before next communication arrives.

\subsection{Scalability and Network Utilization}\label{sec:scalability}
\vspace{-1mm}

\autoref{fig:scale_results}.a shows the total training loop algorithm latency, decomposed into total compute and exposed communication, for the three different workloads, running on different torus network sizes. As \autoref{fig:scale_results}.a shows, the exposed communication delay increases with network size, due to the increased overhead of communication since it requires more steps to be finished. It also shows that among the two baselines with overlap, baselineCompOpt always outperforms BaselineCommOpt The reason is that any saving in compute time directly translates to the reduction in iteration time, while communication may be overlapped with compute and does not impact total training time. However, as we showed in this paper, exposed communication, due to the poor network utilization, can limit the training performance, if left unaddressed. Another interesting point is the BaselineNoOverlap. Compared to the BaselineCommOpt, BaselineNoOverlap is always better thanks to the huge savings in compute time. When comparing to BaselineCompOpt, it works worse for all workloads except for ResNet-50 running on systems larger than 16 NPUs. The reason is combining all small-size communication of ResNet-50 and running them together results in better network performance than running small collectives individually. In all cases, \ncu outperforms all baselines. According to this figure and when averaging across different sizes and workloads, BaselineNoOverlap,  BaselineCommOpt, BaselineCompOpt and \ncu achieves the 68.5$\%$, 49.9$\%$, 75.7$\%$, and 91$\%$ of the ideal system performance (in terms of iteration time), respectively.

\autoref{fig:scale_results}.b shows the performance improvement of \ncu over all baselines as the network size scales. According to \autoref{fig:scale_results}.b, in general, the performance gap of \ncu over the baselines increases by system size. But this increase is more evident in BaselineNoOverlap and BaselineCompOpt since: i) these configurations are more vulnerable to the increased communication overhead, ii) the long compute latency of BaselineCommOpt allows to tolerate more communication overhead for a larger system. Moreover, different workloads get different benefits from \ncu as the system scales since the exposed communication performance is also dependent on the interplay between communication size, intervals between communication tasks, and the compute times to overlap communication with. On average and across different network sizes, when compared to the best baseline at any configuration, \ncu achieves 1.41$\times$ (up to 1.51$\times$), 1.12$\times$ (up to 1.17$\times$), and 1.13$\times$ (up to 1.19$\times$) speedup in iteration time for ResNet-50, GNMT, and DLRM, respectively. \autoref{fig:scale_results}.b also preserves the ratio of the effective network BW utilization (in terms of GB/s) between \ncu and different baselines, since for each specific network size/workload, different configurations inject the same amount of traffic to the network. When averaging across all points in \autoref{fig:scale_results}.b, \ncu can improve the effective network utilization by 1.44$\times$ (up to 2.67) over all baselines.

\subsection{DLRM Optimization}\label{sec:DLRMOptimization}
The extra available memory bandwidth enabled by \ncu offers an opportunity to perform various workload level optimizations to further increase the training performance. Here we pick DLRM as one example for such optimization. It is possible to use extra memory BW to overlap the (memory intensive) embedding lookup/update of the next/previous iteration with the computation of the current iteration. This is because embedding indices are barely reused in the consecutive training iteration \cite{dlrm}. This way, the embedding operation goes outside of the critical path in the training loop. To demonstrate this, we performed a simple experiment where we allocate one SM and 80 GB/s available memory BW for embedding update/lookup of the previous/next iteration. For the next iteration embedding lookup, we immediately issue communication once the lookup is finished and don't wait until the next iteration. \autoref{fig:optimization} shows the impact of such optimization on the baseline vs. \ncu. The total computation time is decreased in both systems since the training loop does not wait for embedding update/lookup at the end/beginning of each iteration. However, BaselineCompOpt benefits little as a result of such optimization due to its poor communication performance. In this case, BaselineCompOpt and \ncu achieve 1.05$\times$ and 1.2$\times$  performance improvement compared to their default training loop, respectively.  
\insertFigureNew{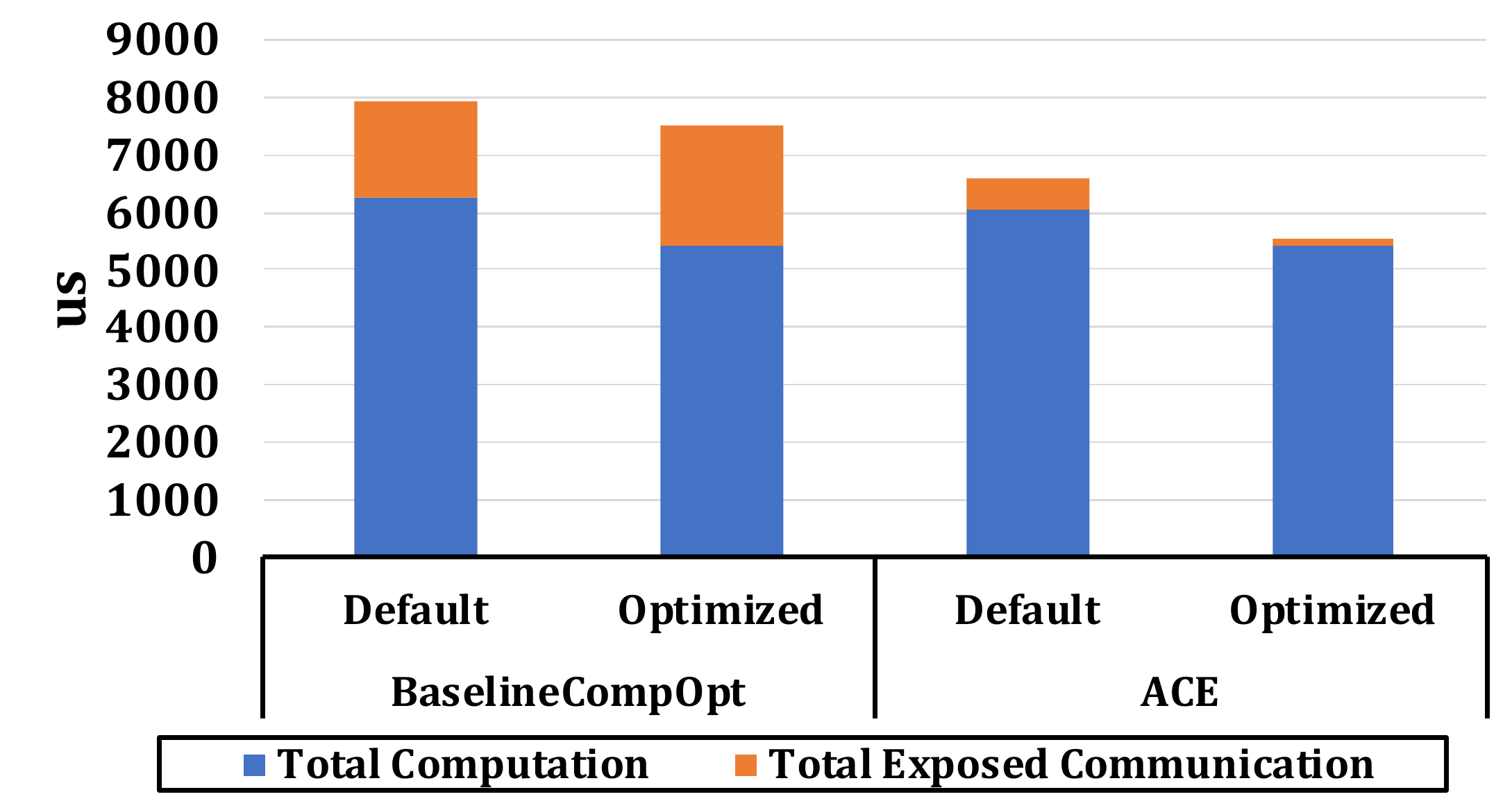}{Impact optimized training loop for the baseline vs. \ncu}{0.75}

\vspace{-2mm}
\section{Related Work}\label{sec:related}
\vspace{-2mm}

\ncu is the first work proposing a dedicated hardware engine at the endpoint for running DL collectives within the Accelerator Fabric.
The works closest in flavor to \ncu 
are those on 
collective offload to datacenter NICs and switches.
We contrast them 
in \autoref{table:related_offload} and discuss 
some key works here.


\textbf{Collective Offload for DL.}
Switch-based offload solutions, such as Intel's Barefoot \cite{Barefoot}, Mellanox SHARP~\cite{Mellanox}, and NVIDIA's shared memory switch offload \cite{InNetwork}, have proposed aggregation in switches.
Switch offload has also been explored for accelerating Reinforcement learning~\cite{CommBottleneck1}.
Unfortunately, as discussed earlier in \autoref{sec:endpoint_vs_switch}, these solutions are restricted to switch-based topologies (limiting their applicability to NVswitch based NVIDIA platforms today). In contrast, \ncu can work with both switch-based and point-to-point topologies that 
are highly popular (e.g., \cite{cloudtpu,HabanaPtP,IntelPtP}) in most training platforms today.

\textbf{Collective Offload in HPC Systems.}
HPC systems with Torus-based topology, such as BlueGene \cite{BlueGene}, PERCS \cite{PERCS}, and Anton2 \cite{Anton}, have supported collective offload on network routers. However, these approaches are designed for 
CPUs and communicate via message passing
rather than accelerators like TPUs/GPUs communicating via 
shared-memory fabrics.
\section{Conclusions}\label{sec:conclusions}
\vspace{-2mm}


In this paper, we identified the issues with compute and memory bandwidth sharing at DL accelerator endpoints in modern distributed training platforms that limit compute-communication overlap.
We made a case for optimizing the endpoint with a novel collectives accelerator called \ncu. We demonstrated that \ncu can efficiently drive the hierarchical AF fabric to close to its peak bandwidth and free up critical compute and memory resources 
for DL computations, speeding up end-to-end training. On average, \ncu frees up the endpoint's required memory BW by 3.5$\times$ to drive the same network BW compared to state-of-the-art baselines. For modern DL workloads and different network sizes, \ncu, on average, increases the effective network bandwidth utilization by 1.44$\times$ (up to 2.67$\times$), resulting in an average of 1.41$\times$ (up to 1.51$\times$), 1.12$\times$ (up to 1.17$\times$), and 1.13$\times$ (up to 1.19$\times$) speedup in iteration time for ResNet-50, GNMT and DLRM when compared to the best baseline configuration, respectively. This work opens up future research in optimizing 
compute-memory-network interactions during distributed training.

\section{Acknowledgment}
This work was supported by awards from Facebook and Intel.
We thank Ching-Hsiang Chu from Facebook for his help with collecting real system measurements in \autoref{sec:motivation}. We also thank Anand Samajdar, from Georgia Tech,  for his help with modifying the compute-simulator and getting the compute times for the simulations in \autoref{sec:results}.
\bibliographystyle{IEEEtranS}
\bibliography{bib/main}

\end{document}